\documentclass[useAMS,usenatbib]{mn2e}

\usepackage{graphicx}
\usepackage{amsmath}
\usepackage{amssymb}
\usepackage{color}
\usepackage{subfig}
\usepackage[breaklinks,colorlinks,citecolor=blue,linkcolor=red]{hyperref} 
\usepackage[all]{hypcap}

\newcommand\msun{\, \rm M_\odot}

\newcommand\eout{{e_{\rm out}}}
\newcommand\eoutsq{{e^2_{\rm out}}}

\newcommand\aout{{a_{\rm out}}}
\newcommand\ain{{a_{\rm in}}}
\newcommand\ein{{e_{\rm in}}}
\newcommand\msmbh{{M_{\rm SMBH}}}
\newcommand\tmerg{{t_{\rm merge}}}

\title[BH and NS mergers in Galactic Nuclei]{Black Hole and Neutron Star Mergers in Galactic Nuclei}
\author[G. Fragione et al.]{  \parbox{\textwidth}{Giacomo Fragione$^{1}$\thanks{E-mail: giacomo.fragione@mail.huji.ac.il}, Evgeni Grishin$^{2}$, Nathan W. C. Leigh$^{3,4,5}$, Hagai B. Perets$^{2}$,\\ Rosalba Perna$^{3,6}$ \vspace*{0.3cm}}\\
$^1$Racah Institute for Physics, The Hebrew University, Jerusalem 91904, Israel\\
$^2$Physics Department, Technion - Israel institute of Technology, Haifa 3200002, Israel\\
$^3$Departamento de Astronom\'ia, Facultad de Ciencias F\'isicas y Matem\'aticas, Universidad de Concepci\'on, Concepci\'on, Chile\\
$^4$Department of Astrophysics, American Museum of Natural History, New York, NY 10024, USA\\
$^5$Department of Physics and Astronomy, Stony Brook University, Stony Brook, NY 11794-3800, USA\\
$^6$Center for Computational Astrophysics, Flatiron Institute,  New York, NY 10010, USA}

\hypersetup{draft}

\begin{document}

\maketitle

\begin{abstract}
Nuclear star clusters surrounding supermassive black holes (SMBHs) in galactic nuclei contain large numbers of stars, black holes (BHs) and neutron stars (NSs), a fraction of which are likely to form binaries. These binaries were suggested to form a triple system with the SMBH, which acts as a perturber and may enhance BH and NS mergers via the Lidov-Kozai mechanism. We follow-up previous studies, but  for the first time perform an extensive statistical study of BH-BH, NS-NS and BH-NS binary mergers by means of direct high-precision regularized $N$-body simulations, including Post-Newtonian (PN) terms up to order PN2.5. We consider different SMBH masses, slopes for the BH mass function, binary semi-major axis and eccentricity distributions, and different spatial distributions for the binaries. We find that the merger rates are a decreasing function of the SMBH mass and are in the ranges $\sim 0.17$-$0.52 \ \mathrm{Gpc}^{-3}\ \mathrm{yr}^{-1}$, $\sim 0.06$-$0.10 \ \mathrm{Gpc}^{-3}\ \mathrm{yr}^{-1}$ and $\sim 0.04$-$0.16 \ \mathrm{Gpc}^{-3}\ \mathrm{yr}^{-1}$ for BH-BH, BH-NS and NS-NS binaries, respectively. However, the rate estimate from this channel remains highly uncertain and depends on the specific assumptions regarding the star-formation history in galactic nuclei and the supply rate of compact objects. We find that $\sim 10\%$--$20\%$ of the mergers enter the LIGO band with eccentricities $\gtrsim 0.1$. We also compare our results to the secular approximation, and show that $N$-body simulations generally predict a larger number of mergers. Finally, these events can also be observable via their electromagnetic counterparts, thus making these compact object mergers especially valuable for cosmological and astrophysical purposes.
\end{abstract}


\section{Introduction}

Recently, the LIGO-Virgo\footnote{\url{https://www.ligo.caltech.edu/}} collaboration detected six sources of gravitational waves (GWs), five from merging BH-BH binaries \citep{abbott16a,abbott16b,abbott17,abbott17a,abbott17b} and one from merging NS-NS binaries \citep{abbott17c}. With ongoing improvements to LIGO and upcoming instruments such as \textit{LISA}\footnote{\url{http://www.et-gw.eu}} and the Einstein Telescope\footnote{\url{https://lisa.nasa.gov}}, hundreds of BH-BH, BH-NS and NS-NS binary sources may be detected within a few years. Thus the modeling of the formation and evolution of BH and NS binaries is crucial for interpreting the signals from all the GW sources we expect to observe.

The origins of BH and NS mergers are actively debated. Several scenarios have been proposed, such as isolated binary evolution in the galactic field \citep{bel16b}, gas-assisted mergers \citep{bart17,sto17,tag18}, triple systems in the field \citep{ant17,sil17,ll18} and dynamically formed in dense clusters \citep{wen03,anm14}, mergers of binaries in galactic nuclei \citep{mill2009,ole2009,antoper12,prod15,ant16,van16,chen17,petr17,chen18,fern18,hamer18,hoan18,rand18} and other dense stellar systems \citep{askar17,baner18,cho18,frak18,rod18,rodlo18,sam18}. Each model predicts different rates (generally of the order of $\sim\ \mathrm{few}$ Gpc$^{-3}$ yr$^{-1}$) and can in principle be distinguished from other channels using the observed mass, eccentricity, spin and redshift distributions \citep[see e.g.][]{olea16}. For instance, dynamically assembled mergers are expected to have a non-negligible probability of appearing eccentric when observed \citep{antoper12,samas18,zevin18}.

Most of the literature pertaining to dynamically-induced mergers focuses on BH and NS binaries forming in globular clusters, while only a few studies have paid attention to the formation of compact-object binaries in the vicinity of a super massive black hole (SMBH) and in nuclear star clusters \citep[e.g.][]{hoan18,leigh16,leigh18}. The pioneering work by \citet{antfab2010} and \citet{antoper12} showed that SMBHs can induce Lidov-Kozai (LK) oscillations on BH-BH and NS-NS binaries orbiting in its vicinity \citep{lid62,koz62}, thus enhancing the probability of merging compact binaries. In this scenario, the eccentricity of the BH/NS binary reaches large values \citep[for a review on LK mechanisms see][]{nao16}, then GW emission drives the binary to merge. While \citet{antoper12} adopted a secular treatment for the equations of motion at the quadruple level of approximation, \citet{hoan18} considered soft binaries and the importance of expansion up to the octuple order. These calculations adopt the secular approximation to study triples, which must satisfy hierarchical conditions \citep{nao16}. In some cases, the inner binary may undergo rapid oscillations in the angular momentum and eccentricity, thus the secular theory is not anymore an adequate description of the three-body equations of motion \citep*{antoper12,antognini14,anm14,luo16,ll18,grish18}. For these cases, direct precise $N$-body simulations, including regularization schemes and Post-Newtonian (PN) terms, are required to follow accurately the orbits of the objects up to the final merger \citep{grish18,fraglei18b,fraglei18}. Recently, \citet{van16} used $N$-body simulations of small ($\sim 300$-$4000$ stars) clusters surrounding $10^3$-$10^4\msun$ black holes to study the effect of LK oscillations in dense environments, while \citet{agu18} used few-body simulations to check the merger rates of BH-BH binaries delivered by infalling star clusters at typical distances of $\sim$ few pc.

\begin{figure} 
\centering
\includegraphics[scale=0.45]{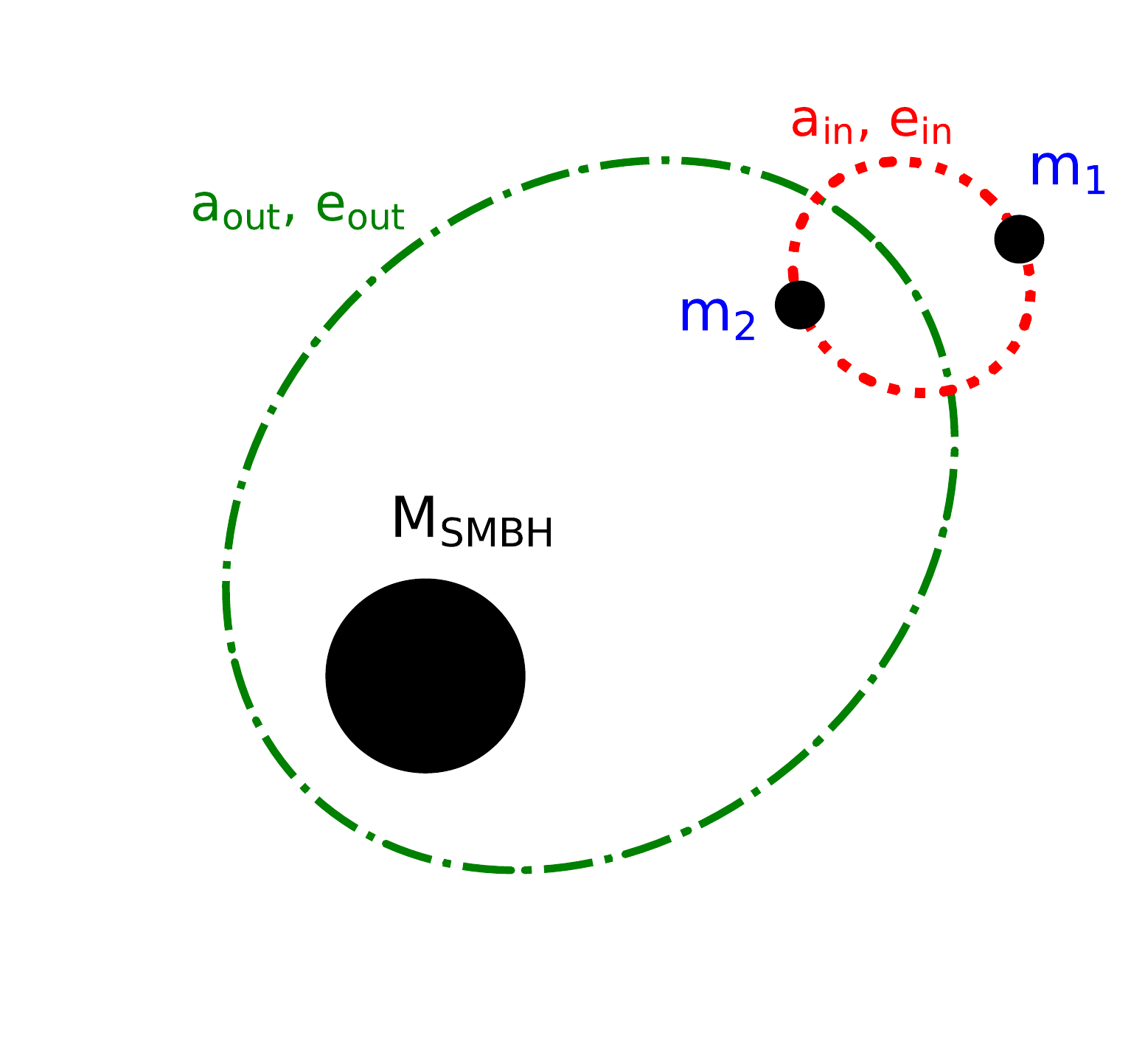}
\caption{The three-body system studied in the present work. We denote the mass of the SMBH as $\msmbh$, and the masses of the binary components as $m_1$ and $m_2$. The semimajor axis and eccentricity of the outer orbit are $\aout$ and $\eout$, respectively, while for the inner orbit they are $\ain$ and $\ein$, respectively.}
\label{fig:threebody}
\end{figure}

In this paper, we revisit the SMBH-induced mergers of compact binaries orbiting in its vicinity. We consider a three-body system consisting of an inner binary comprised of a BH-BH/NS-NS/BH-NS binary, and an outer binary comprised of the SMBH and the centre of mass of the inner binary. Figure~\ref{fig:threebody} depicts the system we study in the present paper. We denote the mass of the SMBH as $\msmbh$ and the mass of the objects in the inner binary as $m_1$ and $m_2$, while the semimajor axis and eccentricity of the inner orbit are $\ain$ and $\ein$, respectively, and for the outer orbit, these are $\aout$ and $\eout$, respectively. While previous papers mainly adopted a secular approximation for the equations of motion, with a few direct N-body integrations in comparison to secular evolution \citep[e.g.][]{antoper12,hoan18}, here, we make the first systematic and statistical study of BH-BH, NS-NS and BH-NS mergers in the proximity of an SMBH by means of direct high-precision $N$-body simulations, including Post-Newtonian (PN) terms up to order PN2.5. Moreover, we consider how different masses of the SMBH affect the mergers of compact binaries, and adopt a mass spectrum for the BHs, while also studying different spatial distributions for the merging binaries. Finally, we discuss observational diagnostics that can help discriminate this compact object merger channel from other ones. 

The paper is organized as follows. In Section~\ref{sect:bhnsnuclei}, we discuss the properties and dynamics of BHs and NSs in galactic nuclei. In Section~\ref{sect:secular}, we discuss the state-of-the-art secular approximations currently being used in the literature.  In Section~\ref{sect:bhnsmergers}, we present our numerical methods to determine the rate of BH-BH, NS-NS and BH-NS mergers in galactic nuclei, for which we discuss the results. In Section~\ref{sect:timedistrates}, we discuss the predicted rate of compact object mergers in galactic nuclei and compare our results to secular approximations, while, in Section~\ref{sect:implications}, we discuss the observational signatures of these events. Finally, in Section~\ref{sect:conc}, we discuss the implications of our findings and draw our conclusions.

\section{Black hole and neutron star binaries in galactic nuclei}
\label{sect:bhnsnuclei}

The evidence in favour of the presence of close binaries composed of compact objects (COs) such as white dwarfs (WDs), and especially NSs and BHs, in dense stellar environments is rapidly growing.  Recently, \citet{hailey17} reported observations of a dozen quiescent X-ray binaries that form a central density cusp within $\sim$ 1 parsec of Sagittarius A$^*$.  The authors argue that the emission spectra they observe are inconsistent with a population of accreting WDs, suggesting that the X-ray binaries must contain mostly NSs and BHs.  However, the relative numbers of these two types of COs is an open question.  Six other X-ray transients are known to be present in the inner parsec of the Galactic Centre, which are also strongly indicative of binaries containing COs \citep[e.g.][]{muno05,hailey17}. Given the very dense stellar environments in galactic nuclei combined with the presence of a central SMBH, these CO binaries can undergo several fates. They can be hardened to shorter orbital periods, either by direct scattering interactions with single stars \citep{per09b,leigh18}, or by LK oscillations due to the SMBH combined with gravitational wave (GW) emission acting at pericentre \citep[e.g.][]{antoper12,prod15}.

In their Table 1, \citet{generosoz18} combine the reported statistics from the literature to provide estimates for the numbers of BHs and NSs in the Galactic Centre. In short, the number of NS X-ray Binaries (XRBs) per stellar mass in the Galactic Centre is roughly three orders of magnitude higher than in the field, and comparable to the number expected to be in globular clusters. Similarly, the number of BH XRBs per stellar mass is roughly three orders of magnitude higher than in the field, and roughly an order of magnitude higher than in any known globular cluster \citep[e.g.][]{strader12,leigh14,leigh16}. 

These large numbers of NS and BH binaries must come from somewhere. Yet, little is known about binaries in our Galactic Centre. A likely explanation is that they are the remnants of massive O/B stars, since hundreds of these are known to be present in the inner $\sim 1$ pc of Sagittarius A*. In the Solar neighborhood, the massive-star binary fraction is very high ($\gtrsim 70\%$) and the most massive binaries have semi-major axes of up to a few AU \citep{sana12}. This alone is suggestive of a high rate of NS and BH formation in nuclear star clusters \citep[e.g.][]{Levin03,genzel08}. Interestingly, the discovery of even a single magnetar within $\lesssim 0.1$ pc of Sgr A* would also argue in favour of a high rate of NS formation, given their short active lifetimes \citep{mori13}. 

This population of CO binaries is continuously depleted through dynamical interactions with other stars and COs (evaporation) and GW mergers. A few mechanisms may replenish the CO population. The binaries may come from outside the central region near the SMBH, which thus serves as a continuous source term \citep{hop09,alex17}. In this scenario, CO binaries form far from the innermost region around the SMBH and gradually migrate on a 2-body relaxation timescale,
\begin{eqnarray}
T_{2b}&=&1.6\times 10^{10}\mathrm{yr}\ \left(\frac{\sigma}{300\ \mathrm{km s}^{-1}}\right)^3\left(\frac{m}{\msun}\right)^{-1}\times\nonumber\\
&\times&\left(\frac{\rho}{2.1\times 10^{6}\msun\ \mathrm{pc}^{-3}}\right)^{-1}\left(\frac{\ln \Lambda}{15}\right)^{-1} \ ,
\label{eqn:t2b}
\end{eqnarray}
towards smaller distances, where they become active in the Lidov-Kozai regime. Here, $\rho$ and $\sigma$ are the 1-D density and velocity dispersion in the Galactic Centre, respectively, $\ln \Lambda$ is the Coulomb logarithm and $m$ is the average stellar mass. On the other hand, our Galactic Centre contains a large population of young massive O-type stars, many of which have been observed to reside in a stellar disk. Likely, most of them were born \textit{in-situ} as a consequence of the fragmentation of a gaseous disk formed from an infalling gaseous clump \citep{genz10}. An important question is which process can make their orbit, which are not observed closer than $\sim 0.05$ pc, approach the SMBH, where efficient Lidov-Kozai oscillations take place. Both planet-like migration in the gaseous disc \citep{baru11} and disc instability \citep{madi09} have been proposed to make the \textit{in situ} binaries migrate, but to which innermost distance with respect to the SMBH is not known exactly. Other mechanisms include triple/quadruple disruptions, where a triple/quadruple is disrupted and the inner binary is left orbiting the SMBH \citep{per09,Gin11,fgu18,fgi18}, and infalls of star clusters \citep*{antmer,fck17}.

Outside our own Milky Way, \citet{secunda18} recently showed that CO binaries can form efficiently in Active Galactic Nucleus (AGN) disks. The COs migrate in the disk due to differential torques exerted by the gas, moving toward migration traps, where the torques actually cancel \citep[e.g.][]{bellovary16}. The COs drift toward the trap and get stuck there, waiting for the next CO to migrate toward it.  Once close enough, the two COs can undergo a strong interaction and end up forming a bound binary due to the dissipative effects of the gas.  COs can also accrete from the disk gas in this scenario, possibly growing substantially and in some cases even forming an IMBH \citep{mckernan12,mckernan14}.  

\begin{table*}
\caption{Models: name, SMBH mass ($M_\mathrm{SMBH}$), binary type, slope of the BH mass function ($\beta$), slope of the outer semi-major axis distribution ($\alpha$), $\ain$ distribution, $\ein$ distribution, merger fraction ($f_{\rm merge}$).}
\centering
\begin{tabular}{lccccccc}
\hline
Name &	$M_\mathrm{SMBH}$ (M$_\odot$) & Binary Type & $\beta$ & $\alpha$ & $f(\ain)$ & $f(\ein)$ & $f_{\rm merge}$ \\
\hline\hline
MW	& $4\times 10^6$ & BH-BH & $1$ & $2$   & \citet{hoan18}    & uniform &  $0.045$ \\
MW	& $4\times 10^6$ & BH-BH & $1$ & $2$   & \citet{antoper12} & thermal & $0.261$ \\
MW	& $4\times 10^6$ & BH-BH & $2$ & $2$   & \citet{hoan18}    & uniform & $0.05$ \\
MW	& $4\times 10^6$ & BH-BH & $3$ & $2$   & \citet{hoan18}    & uniform & $0.036$ \\
MW	& $4\times 10^6$ & BH-BH & $4$ & $2$   & \citet{hoan18}    & uniform & $0.041$ \\
MW	& $4\times 10^6$ & BH-BH & $1$ & $0$   & \citet{hoan18}    & uniform & $0.035$ \\
MW	& $4\times 10^6$ & BH-BH & $1$ & $3$   & \citet{hoan18}    & uniform & $0.051$ \\
MW	& $4\times 10^6$ & BH-NS & $1$ & $2$   & \citet{hoan18}    & uniform & $0.026$ \\
MW	& $4\times 10^6$ & BH-NS & $2$ & $2$   & \citet{hoan18}    & uniform & $0.029$ \\
MW	& $4\times 10^6$ & BH-NS & $3$ & $2$   & \citet{hoan18}    & uniform & $0.028$ \\
MW	& $4\times 10^6$ & NS-NS & $-$ & $1.5$ & \citet{hoan18}    & uniform & $0.032$ \\
MW	& $4\times 10^6$ & NS-NS & $-$ & $2$   & \citet{hoan18}    & uniform & $0.028$ \\
MW	& $4\times 10^6$ & NS-NS & $-$ & $2$   & \citet{antoper12} & thermal & $0.079$ \\
\hline
GN	& $1\times 10^8$ & BH-BH & $1$ & $2$   & \citet{hoan18}    & uniform & $0.072$ \\
GN	& $1\times 10^8$ & BH-BH & $1$ & $2$   & \citet{antoper12} & thermal & $0.258$ \\
GN	& $1\times 10^8$ & BH-NS & $1$ & $2$   & \citet{hoan18}    & uniform & $0.054$ \\
GN	& $1\times 10^8$ & NS-NS & $-$ & $2$   & \citet{hoan18}    & uniform & $0.044$ \\
GN	& $1\times 10^8$ & NS-NS & $-$ & $2$   & \citet{antoper12} & thermal & $0.133$ \\
\hline
GN2	& $1\times 10^9$ & BH-BH & $1$ & $2$   & \citet{hoan18}    & uniform & $0.087$ \\
GN2	& $1\times 10^9$ & BH-BH & $1$ & $2$   & \citet{antoper12} & thermal & $0.3$ \\
GN2	& $1\times 10^9$ & BH-NS & $1$ & $2$   & \citet{hoan18}    & uniform & $0.061$ \\
GN2	& $1\times 10^9$ & NS-NS & $-$ & $2$   & \citet{hoan18}    & uniform & $0.048$ \\
GN2	& $1\times 10^9$ & NS-NS & $-$ & $2$   & \citet{antoper12} & thermal & $0.055$ \\
\hline
\end{tabular}
\label{tab:models}
\end{table*}

\subsection{Outer semi-major axis and eccentricity}
 
The numbers and spatial profiles of BHs and NSs are poorly known in galactic nuclei. In general, stars tend to form a power-law density cusp around an SMBH. The classical result by \citet{bahcall76} shows that a population of equal-mass objects forms a power-law density cusp around an SMBH, $n(r)\propto r^{-\alpha}$, where $\alpha=7/4$. For multi-mass distributions, lighter and heavier objects develop shallower and steeper cusps, respectively \citep{hopm06,Fre06,ale09,pre10,aharon16,baumg18}, while source terms may make the cusp steeper as well \citep{aharon15,frasar18}. Recent observations of the Milky Way's centre showed that the slope of the cusp appears to be shallower \citep[$\alpha\sim 5/4$;][]{gall18,scho18}. As BHs and NSs are heavier than average stars, they are expected to relax into steeper cusps, with BHs relaxing into steeper cusps than NSs \citep{Fre06,hopal06}. If not all BHs have  the same mass, but follow a mass distribution, only the more massive ones would have steeper slopes, while low-mass BHs should follow shallower slopes \citep{aharon16}.

In the present study, we assume that the BH and NS number densities follow a cusp with $\alpha=2$. We also study the effects of the cusp slope, by considering a steeper cusp ($\alpha=3$) and a uniform density profile ($\alpha=0$) for BHs, and a shallower cusp ($\alpha=1.5$) for NSs. For the maximum outer semi-major axis, we take $a_{\rm out}^{M}=0.1$ pc following \citet{hoan18}, which approximately corresponds to the value at which the eccentric Lidov-Kozai timescale is equal to the timescale over which accumulated fly-bys from single stars tend to unbind the binary (see Eqs.~\ref{eqn:binevap}-\ref{t_oct}). However, the ratio of these two timescales generally depends on the binary and cusp properties since wide binaries could be affected by the LK mechanism at larger distances. We also consider one model where we take $a_{\rm out}^{M}=0.5$ pc, to assess how our results depend on this parameter. As discussed, \textit{in-situ} formation occurs in our Galactic Centre at larger distances and some mechanism that delivers such CO binaries closer to the SMBH has to be invoked. Finally, we sample the outer orbital eccentricity from a thermal distribution \citep{jeans1919}.

\subsection{Inner semi-major axis and eccentricity}

The inner binary (BH-BH/NS-NS/BH-NS) semi-major axis and eccentricity are not well known. Different models predict different distributions. Moreover, the dense environments characteristic of galactic nuclei should cause both distributions to diffuse over time, thus changing the relative distributions. \citet{hop09} made the only attempt to model binaries very close to the SMBH, even though this pioneering study accounted only for the 2-body relaxation process. Other relaxation processes, such as resonant relaxation \citep{rauch96}, may affect the distribution as well \citep{hamer18}. For what concerns eccentricity, though it mostly depends on the natal-kick and the common-envelope phase, it also depends on the scattering of the CO binaries by local COs and stars. As a consequence, \textit{in-situ} formation would probably favour circular binaries, while the migration scenario would most likely prefer a thermal distribution, as a consequence of the energy exchange through many dynamical encounters of the CO binaries with other stars and COs in galactic nuclei.

\citet[][see Fig.~1]{antoper12} used the results of \citet{belc04} for the initial distribution of BH-BH binaries orbits, while, for NSs, they used the observed pulsar binary population as found in the ATNF pulsar catalog \footnote{http://www.atnf.csiro.au/research/pulsar/psrcat} \citep{manc05}. We note that their distributions referred to isolated binaries and used a simplified approach for to account for the softening and binary-destruction due to the crowded environments (following the approach in \citep{per07}). Inner eccentricities were sampled from a uniform distribution. Recently, \citet{hoan18} drew the inner semi-major axes from a log-uniform distribution in the range $0.1$-$50$ AU, somewhat consistent with the observed distribution from \citet{sana12}, which favors short period binaries, and the inner eccentricities from a uniform distribution \citep{ragh10}. Note the caveat that this distribution corresponds to massive MS binaries, and not to CO binaries which already evolved and change their configurations. 

\subsection{Masses}
The mass distribution of BHs is unknown, even in isolation. Moreover, in galactic nuclei, irrespective of the original mass function, mass-segregation can make the effective BH mass function even steeper in the inner-most regions \citep{aharon16}. This would in turn also affect the outer orbit distribution, since more massive BHs would have steeper slopes \citep{aharon16}.

In our models, we sample the masses of the BHs from
\begin{equation}
\frac{dN}{dm} \propto M^{-\beta}\ ,
\label{eqn:bhmassfunc}
\end{equation}
in the mass range $5\msun$-$100\msun$\footnote{Note that pulsational pair instability may limit the maximum mass to $\sim 50\msun$ \citep{bel2016}.}. To check how the results depend on the slope of the BH mass function, we run models with $\beta=1$, $2$, $3$, $4$ for both the BH-BH and BH-NS binaries \citep{olea16}. For NSs, we fix the mass to $1.3\msun$ \citep*[e.g.][]{fpb18}.

\subsection{Inclinations and relevant angles}
We draw the initial mutual inclination $i_0$ between the inner and outer orbit from an isotropic distribution (i.e. uniform in $\cos i$). The other relevant angles, such as the arguments of pericentre, nodes and mean anomalies, are drawn randomly.

\subsection{Timescales in galactic nuclei}
In the dense stellar environment of a galactic nucleus, several dynamical processes other than 2-body relaxation (Eq.~\ref{eqn:t2b}) can take place and affect the evolution of the stellar and compact object populations. On smaller timescales than $T_{2b}$, resonant relaxation \citep{rauch96,kocs15} randomizes the direction and magnitude (hence eccentricity) of the outer orbit on a typical timescale
\begin{equation}
T_{\rm RR}=9.2\times 10^{8}\ \mathrm{yr}\ \left(\frac{\msmbh}{4\times 10^6 \msun}\right)^{1/2}\left(\frac{a_{out}}{0.1\ \mathrm{pc}}\right)^{3/2}\left(\frac{m}{\msun}\right)^{-1}\ .
\label{eqn:trr}
\end{equation}
On even shorter timescales, vector resonant relaxation changes the direction (hence the relative inclination) of the outer orbit angular momentum on a typical timescale
\begin{eqnarray}
T_{\rm VRR}&=&7.6\times 10^{6}\ \mathrm{yr}\ \left(\frac{\msmbh}{4\times 10^6 \msun}\right)^{1/2}\times\nonumber\\
&\times & \left(\frac{a_{\rm out}}{0.1\ \mathrm{pc}}\right)^{3/2}\left(\frac{m}{\msun}\right)^{-1}\left(\frac{N}{6000}\right)^{-1/2}\ ,
\label{eqn:tvrr}
\end{eqnarray}
where $N$ is the number of stars within $a_{\rm out}$. In the context of Kozai-Lidov oscillations, vector resonant relaxation plays a role, since it may affect the initial inclination of the inner and outer orbit of the CO binary on timescales comparable to or even shorter than the Kozai-Lidov timescale \citep{hamer18}.

Finally, binaries may evaporate due to dynamical interactions with field stars in the dense environment of a galactic nucleus when 
\begin{equation}
\frac{E_b}{(m_1+m_2)\sigma^2}\lesssim 1\ ,
\end{equation}
where $E_b$ is the binary internal orbital energy and $\sigma$ is the velocity dispersion. This happens on an evaporation timescale \citep{binntrem87}
\begin{eqnarray}
T_{\rm EV}&=&3.2\times 10^{7}\ \mathrm{yr} \left(\frac{m_1+m_2}{2\msun}\right)\left(\frac{\sigma}{300\ \mathrm{km s}^{-1}}\right)\left(\frac{m}{\msun}\right)^{-1}\times\nonumber\\
&\times&\left(\frac{a_{\rm in}}{1\ \mathrm{AU}}\right)^{-1}\left(\frac{\rho}{2.1\times 10^{6}\msun\ \mathrm{pc}^{-3}}\right)^{-1}\left(\frac{\ln \Lambda}{15}\right)^{-1}\ .
\label{eqn:binevap}
\end{eqnarray}

\section{Secular averaging techniques}
\label{sect:secular}
The merger time of an isolated binary of component masses $m_1$, $m_2$, semimajor axis $a$ and eccentricity $e$ emitting GWs is \citep{pet64}
\begin{equation}
T_{{\rm GW}}(a,e)=\frac{5}{256}\frac{c^{5}a^{4}}{G^{3}m_{1}m_{2}(m_{1}+m_{2})}(1-e^{2})^{7/2}. \label{eq:t_merge}
\end{equation}
For a triple system made up of an inner binary that is orbited by an outer companion, the inner eccentricity can be pumped by the tidal potential of a distant body via the Lidov-Kozai (LK) mechanism \citep{lid62,koz62}. The LK oscillations occur on a secular timescale \citep{antognini15}
\begin{equation}
t_{\rm sec} = \frac{8}{15\pi}\frac{m_{{\rm tot}}}{m_{{\rm out}}}\frac{P_{{\rm out}}^{2}}{P_{{\rm in}}}(1-\eoutsq)^{3/2}\ ,
\label{t_sec}
\end{equation}
where $m_{\rm out} = \msmbh$ and $m_{\rm tot} = \msmbh + m_{\rm bin} \approx \msmbh $, $P_{{\rm in}}$ and $P_{{\rm out}}$ are the orbital periods of the inner and outer binary, respectively. The large values attained by the inner eccentricity make the overall merger time of the inner binary shorter since it efficiently dissipates energy when $e \sim e_{\rm max}$ \citep[e.g., see][]{antoper12}. The LK-induced merger time is \citep{antoper12,rand18, ll18}
\begin{equation}
T_{\rm GW}^{\rm LK}(a,e_{\rm max}) \approx T_{\rm GW}(a,e_{\rm max})/\sqrt{1-e_{\rm max}^2}\ \label{eq:T_LKGW} .
\end{equation}
The maximal eccentricity is a function mostly of the initial mutual inclination, $i_0$, and is usually evaluated by the secular approximation, which relies on double-averaging of both the inner and the outer orbits \citep{rand18}. In the leading, quadrupole order, the system is integrable and has been widely studied \citep[see recent review by][and references therein]{nao16}, where coupled oscillations between the eccentricity and inclination of the inner binary are excited for sufficiently large initial mutual inclinations. The inner binary eccentricity approaches almost unity as $i_0$ approaches $\sim 90$ deg.

When the outer orbit is eccentric and the inner binary has an extreme mass ratio, octupole-level perturbations turn the system from integrable to chaotic \citep{ln14}, and can potentially induce extreme orbital eccentricities, orbital flips and even direct collisions \citep{Katz11,ln11}. The strength of the octupole perturbation is encapsulated in the octupole parameter as
\begin{equation}
\epsilon_{\rm oct} \equiv \frac{m_1 - m_2}{m_1 + m_2}\frac{\ain}{\aout}\frac{\eout}{1 -\eoutsq}. \label{eps_oct}
\end{equation}
Generally, increasing $\epsilon_{\rm oct}$ will increase the parameter space corresponding to orbital flips and very large eccentricities. The typical timescale for an orbital flip is \citep{antognini15}
\begin{equation}
t_{\rm flip} = \frac{8}{\pi}\sqrt{\frac{10}{\epsilon_{\rm oct}}} t_{\rm sec} \label{t_oct}\,.
\end{equation}

The secular approximation assumes that the triple system is hierarchical, namely that $P_{\rm in} / P_{\rm out} \propto (\ain / \aout)^{2/3} \ll 1$, thus ignoring short-term variations ($t \sim P_{\rm out }$) of the osculating elements, whose typical strength can be parametrized by the so-called 'single-averaging parameter' \citep{luo16}
\begin{equation}
\epsilon_{{\rm SA}}\equiv\left(\frac{\ain}{\aout (1-\eoutsq)} \right)^{3/2}\left(\frac{\msmbh}{m_{\rm bin}}\right)^{1/2}=\frac{P_{{\rm out}}}{2\pi\tau_{{\rm sec}}}.\label{eq:epssa}
\end{equation}
Also, \citet{luo16} point out that in addition to the fluctuating terms, additional secular evolution can take place.  Consequently, the resulting fate of the system could be different, since additional extra apsidal and nodal precession changes the structure of the LK resonance. \citet{grish17} showed that extra apsidal precession shifts the critical inclination for the LK resonance and affects the Hill stability limit of irregular satellites.

Recently, \citet{grish18} used \citet{luo16}'s result to find an analytic formula for the maximal eccentricity that can be reached due to LK oscillations
\begin{align}
e_{\rm max}  & = \bar{e}_{\rm max} + \delta e\ ,\nonumber \\
\bar{e}_{\rm max} & = \sqrt{1-\frac{5}{3}\cos^{2}i_{0}\frac{1+\frac{9}{8}\epsilon_{{\rm SA}}\cos i_{0}}{1-\frac{9}{8}\epsilon_{{\rm SA}}\cos i_{0}}}\ , \nonumber \\
\delta e & =\frac{135}{128}\bar{e}_{\rm max}^{\rm SA}\epsilon_{\rm SA}\left[ \frac{16}{9}\sqrt{\frac{3}{5}}\sqrt{1-\bar{e}_{\rm max}^{2}}+\epsilon_{{\rm SA}}-2\epsilon_{{\rm SA}}\bar{e}_{\rm max}^{2}\right]\ \label{eq:emax_corr}.
\end{align}
If the inner binary is too compact or the tertiary is too far away, the maximal eccentricity could be quenched, e.g. by GR (general relativistic) precession, and the average maximal eccentricity is given by solving \citep{grish18}
\begin{align}
\bar{A}(1-\bar{e}_{{\rm max}}^{2}) & =8\frac{\epsilon_{{\rm GR}}}{\bar{e}_{{\rm max}}^{2}}\sqrt{1-\bar{e}_{\rm max}^{2}} + 15\bar{j}_{z}^{2}\left(1+\frac{9}{8}\epsilon_{{\rm SA}}\bar{j}_{z}\right)\nonumber \\
\bar{A}(\bar{j}_{z},\bar{e}_{{\rm max}}) & \equiv9-\epsilon_{{\rm SA}}\frac{81}{8}\bar{j}_{z}+8\frac{\epsilon_{{\rm GR}}}{\bar{e}_{{\rm max}}^{2}}\,, \label{eq:jmin_e0}
\end{align}
where $\bar{j}_{z} = \sqrt{(1-e_0^2)\cos i_0}$ is the initial normalized angular momentum of the inner binary and
\begin{equation}
\epsilon_{{\rm GR}}\equiv\frac{3m_{{\rm bin}}(1-e_{{\rm out}}^{2})^{3/2}}{m_{{\rm out}}}\left(\frac{a_{{\rm out}}}{a}\right)^{3}\frac{G m_{\rm bin}}{ac^2}\label{eq:epsgr}
\end{equation}
measures the ratio between the apsidal precession rates induced by Lidov-Kozai and GR perturbations. The above formulae are valid in the limit $\epsilon_{\rm oct}=0$. 

In some configurations the inner binary may undergo rapid oscillations in the angular momentum and eccentricity, thus the secular theory is no longer an adequate description of the three-body equations of motion \citep{antoper12,antognini14}. For instance, this can happen when the typical time-scale for the angular momentum of the inner orbit to change by of order itself becomes comparable to (or even shorter than) the outer or inner orbital periods \citep{anm14}. Thus, in this case, the secular approximation can fail to predict both the correct maximum eccentricity and merger time. Although computationally expensive (in particular in the case of a third very massive companion as in this paper), direct $N$-body simulations including Post-Newtonian (PN) terms represent the most reliable option for accurately studying the effects of the tertiary companion in reducing the GW merger time of the inner binary.

\section{N-Body Simulations: black hole and neutron star mergers in galactic nuclei}
\label{sect:bhnsmergers}

\begin{figure} 
\centering
\includegraphics[scale=0.525]{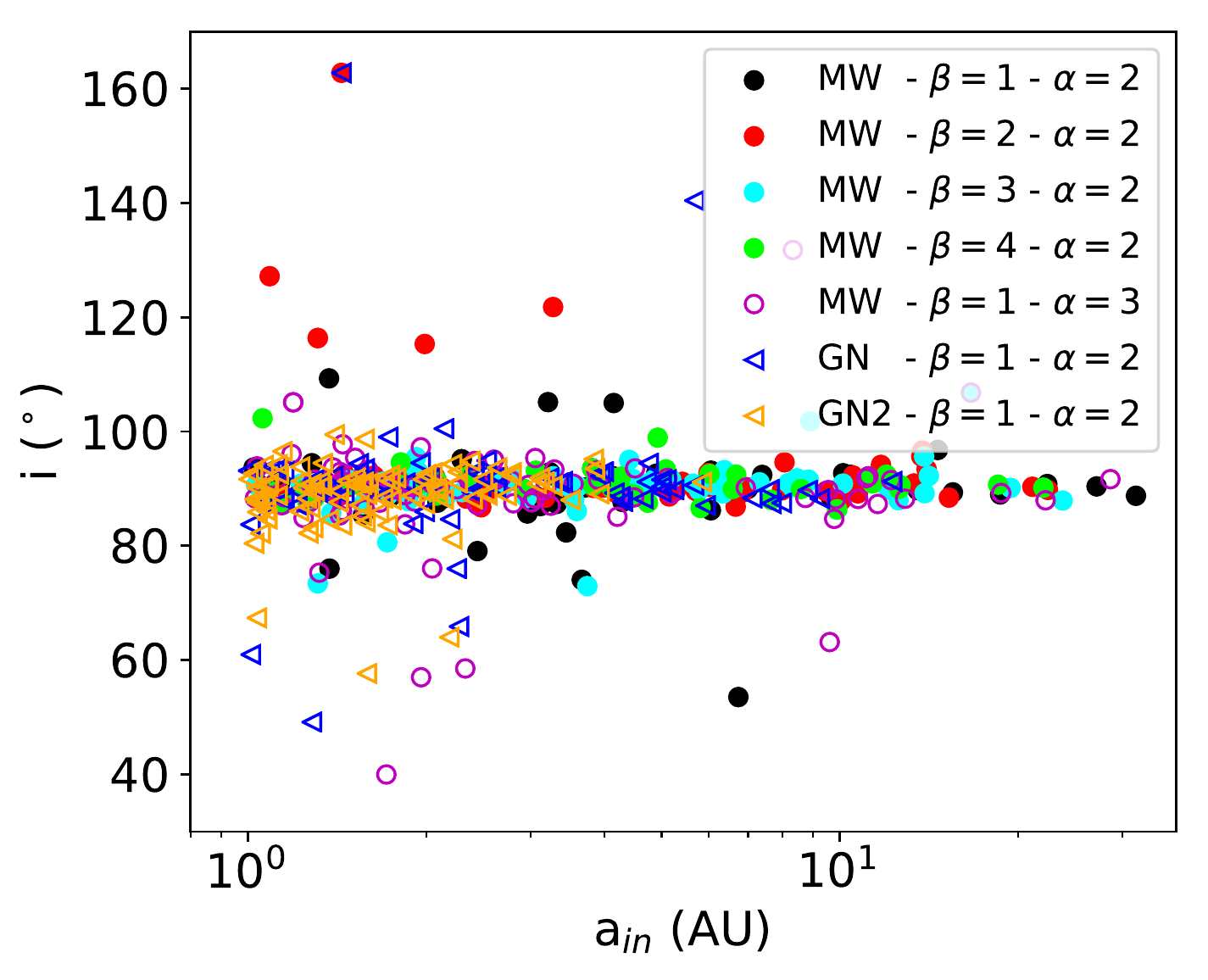}
\includegraphics[scale=0.525]{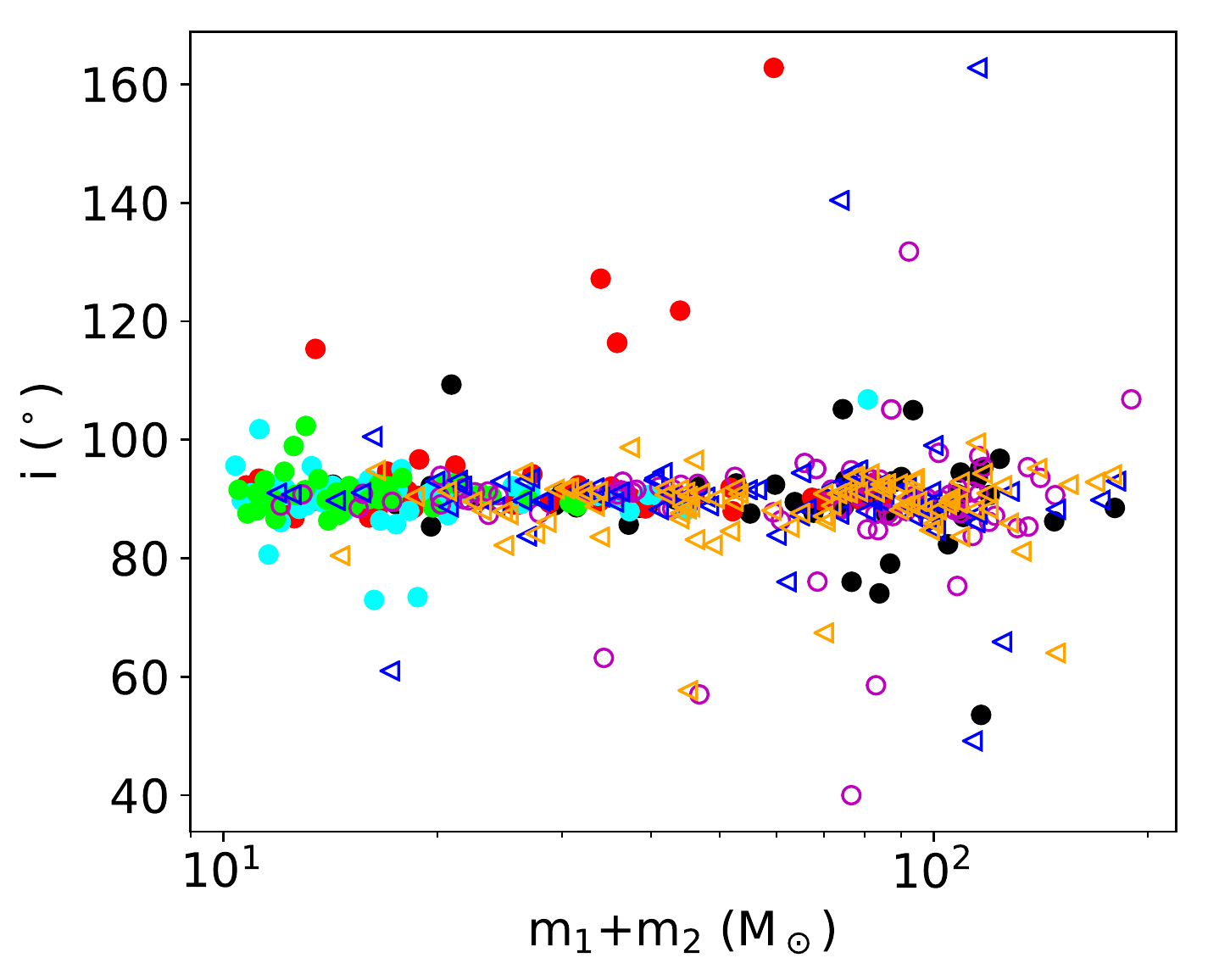}
\caption{Inclination as a function of the initial semi-major axis $\ain$ and the total mass in merged BH-BH binaries for all models with $f(\ain)$ from \citet{hoan18}. Most BH binaries that merge have initial inclinations $\sim 90^\circ$, where the enhancement in the maximum eccentricity is expected to be larger due to Lidov-Kozai oscillations.}
\label{fig:incl}
\end{figure}

\begin{figure*} 
\centering
\begin{minipage}{20.5cm}
\hspace{0.5cm}
\includegraphics[scale=0.5]{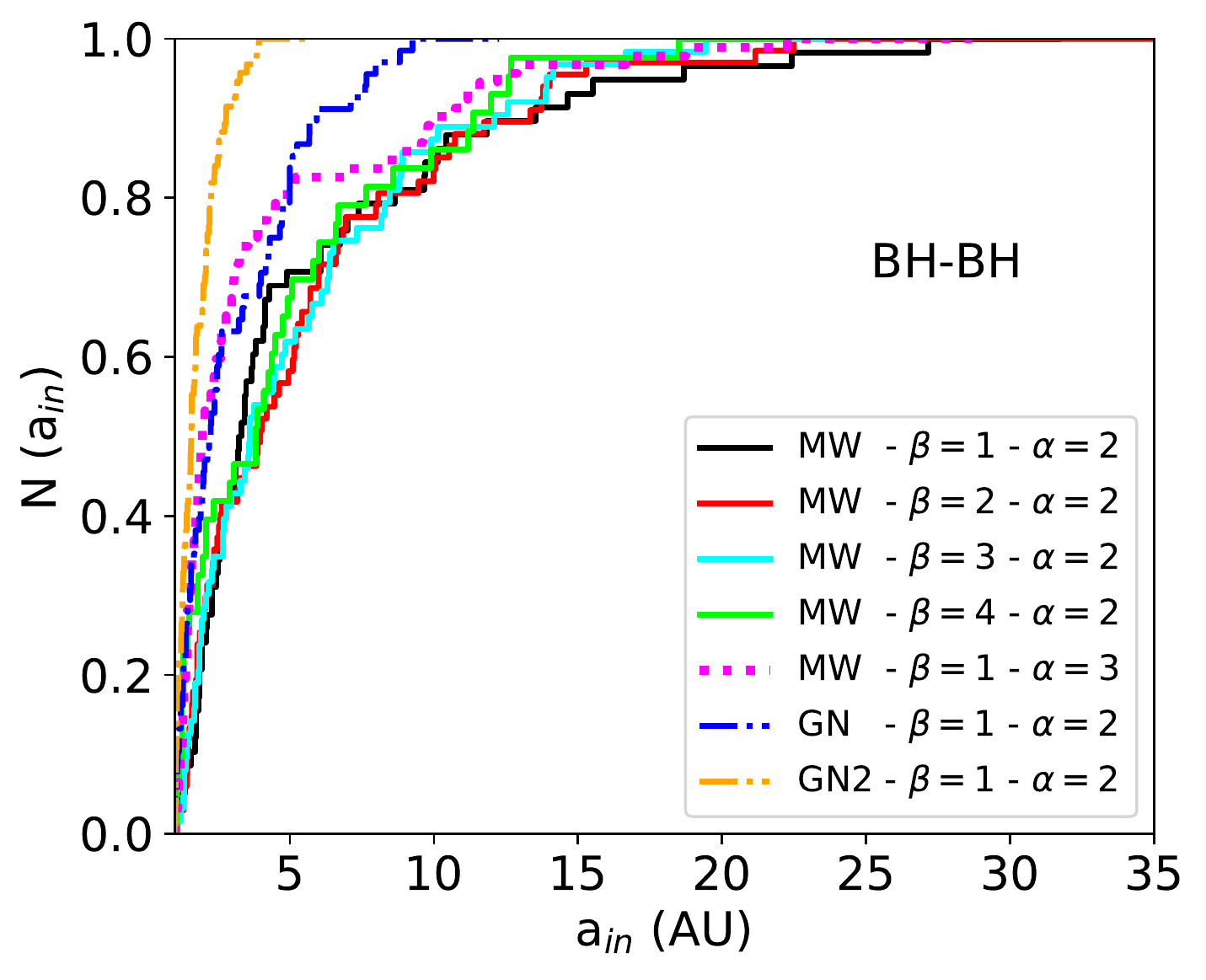}
\hspace{1.5cm}
\includegraphics[scale=0.5]{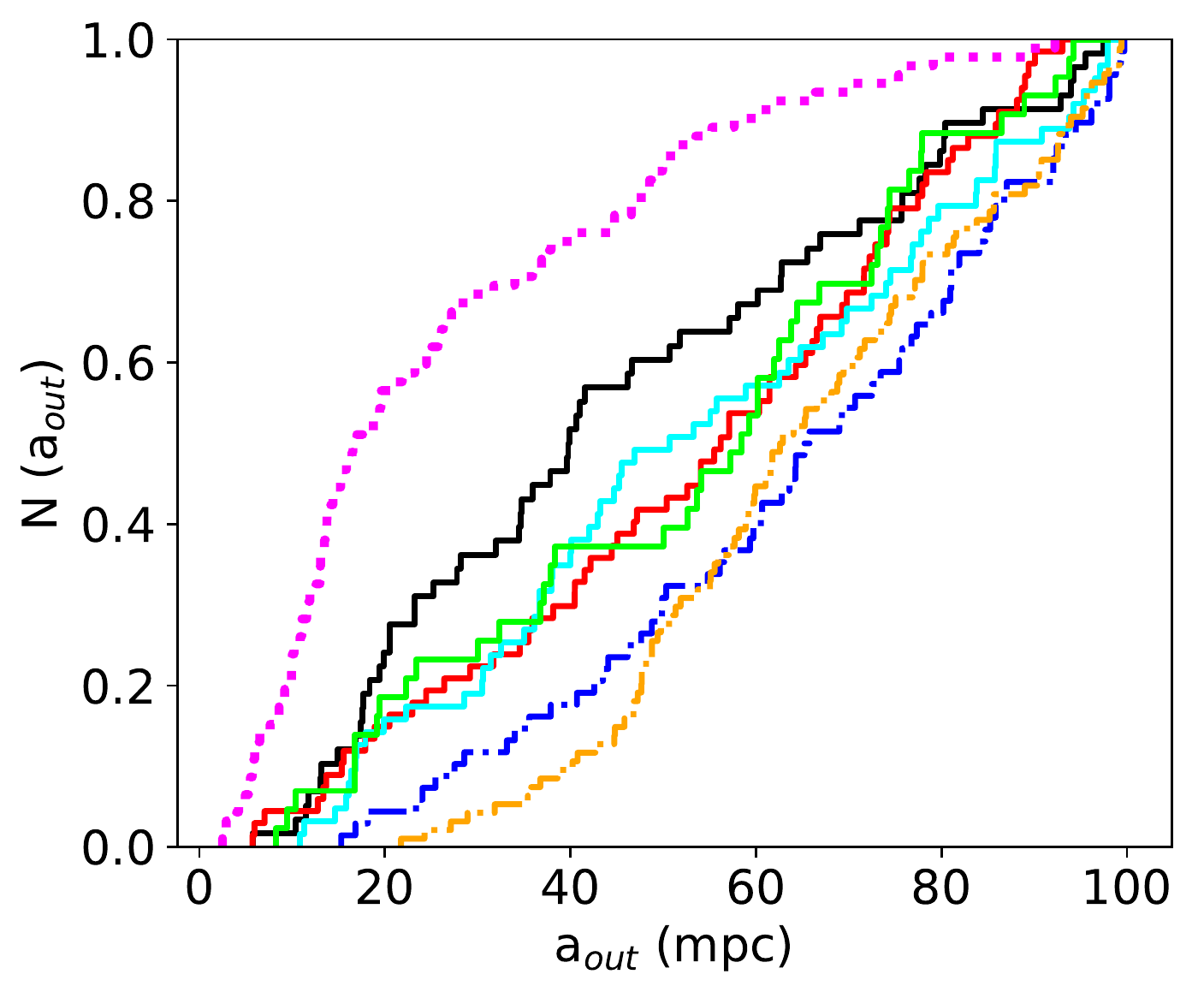}
\end{minipage}
\begin{minipage}{20.5cm}
\hspace{0.5cm}
\includegraphics[scale=0.5]{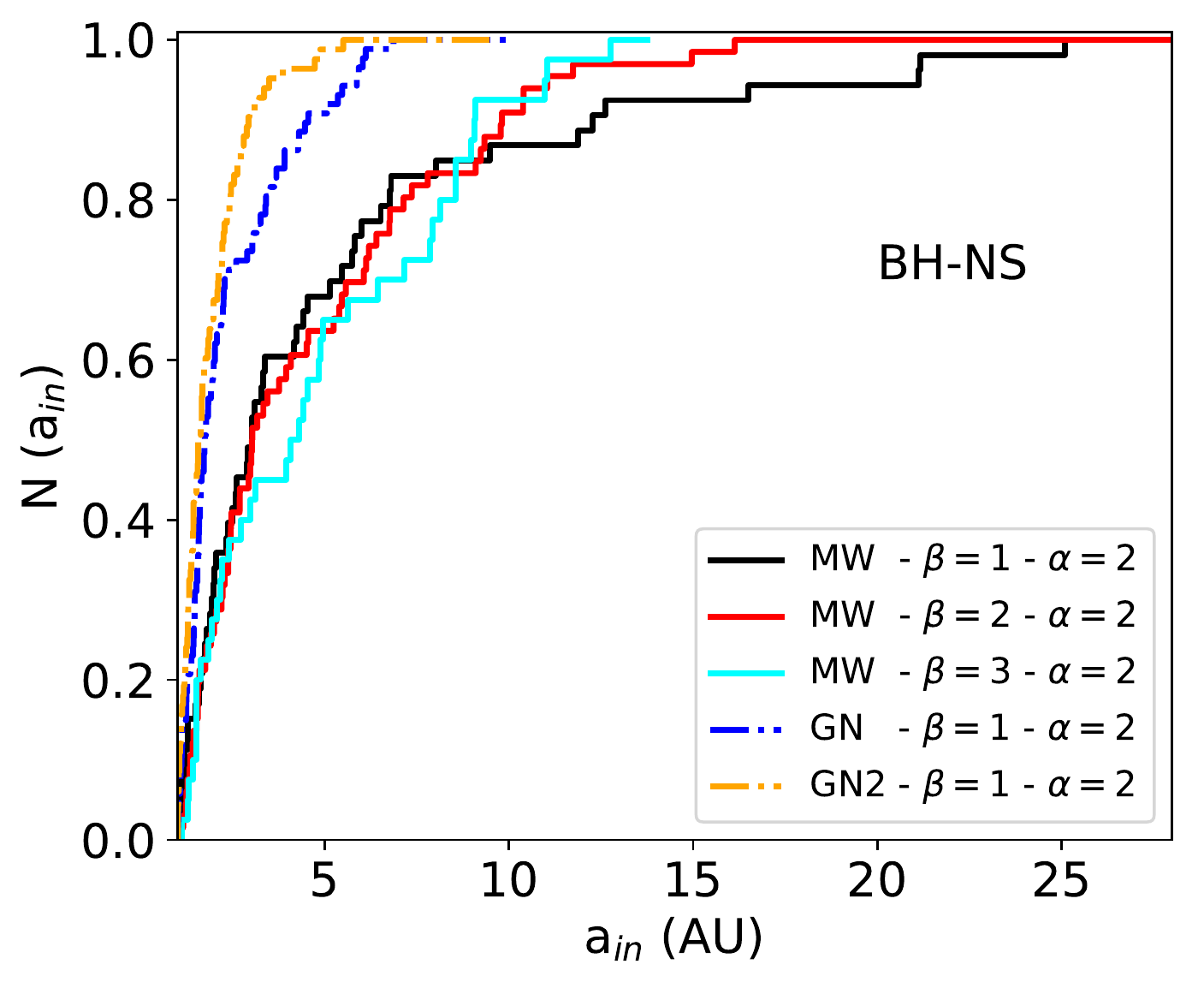}
\hspace{1.5cm}
\includegraphics[scale=0.5]{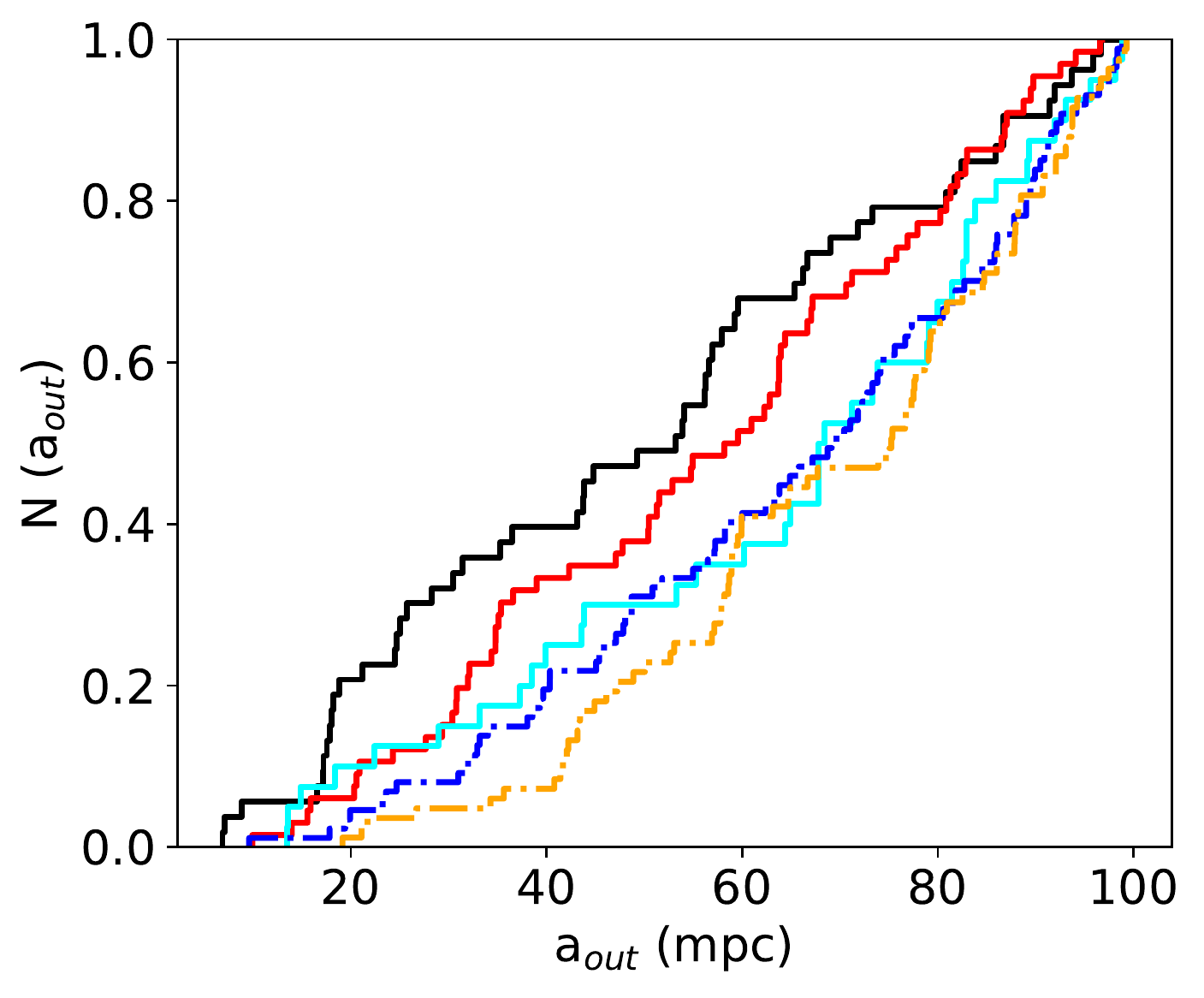}
\end{minipage}
\begin{minipage}{20.5cm}
\hspace{0.5cm}
\includegraphics[scale=0.5]{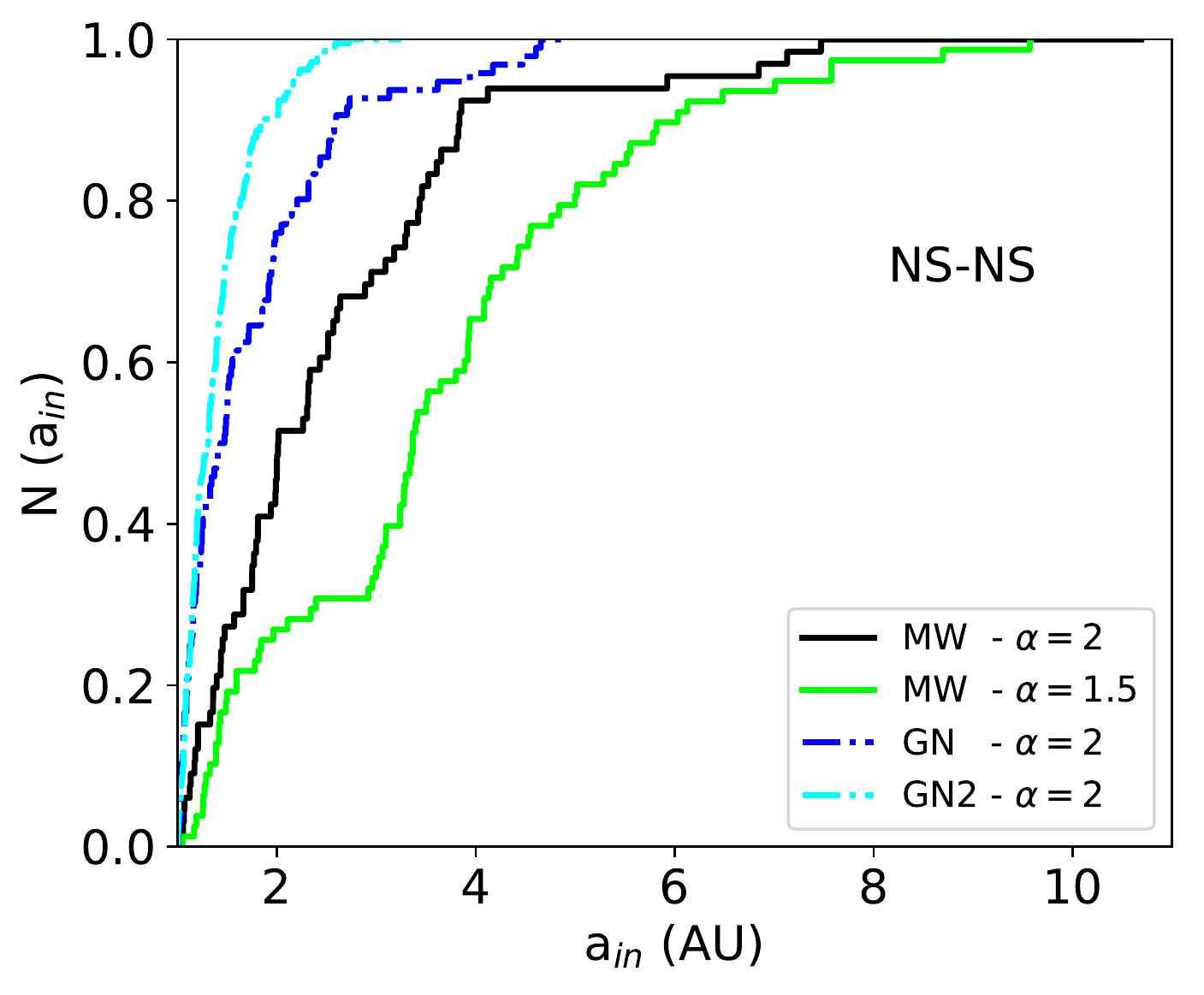}
\hspace{1.5cm}
\includegraphics[scale=0.5]{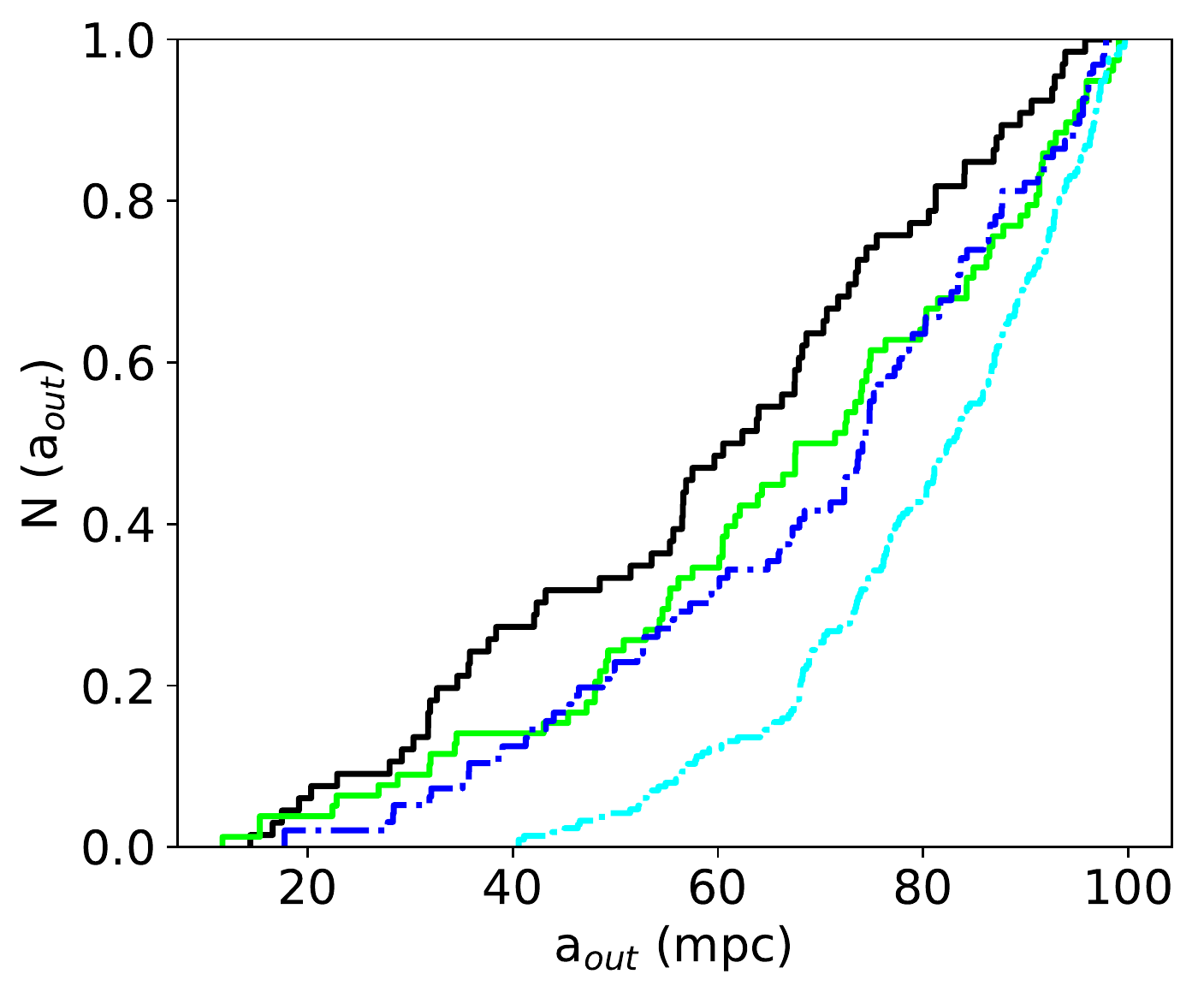}
\end{minipage}
\caption{Cumulative distributions of the inner semi-major axis $\ain$ (left) and of the outer semi-major axis $\aout$ (right) for BH-BH (top), BH-NS (centre), NS-NS (bottom) binaries that merge in all models with $f(\ain)$ from \citet{hoan18}.}
\label{fig:ainaout}
\end{figure*}

\begin{figure*} 
\centering
\begin{minipage}{20.5cm}
\includegraphics[scale=0.55]{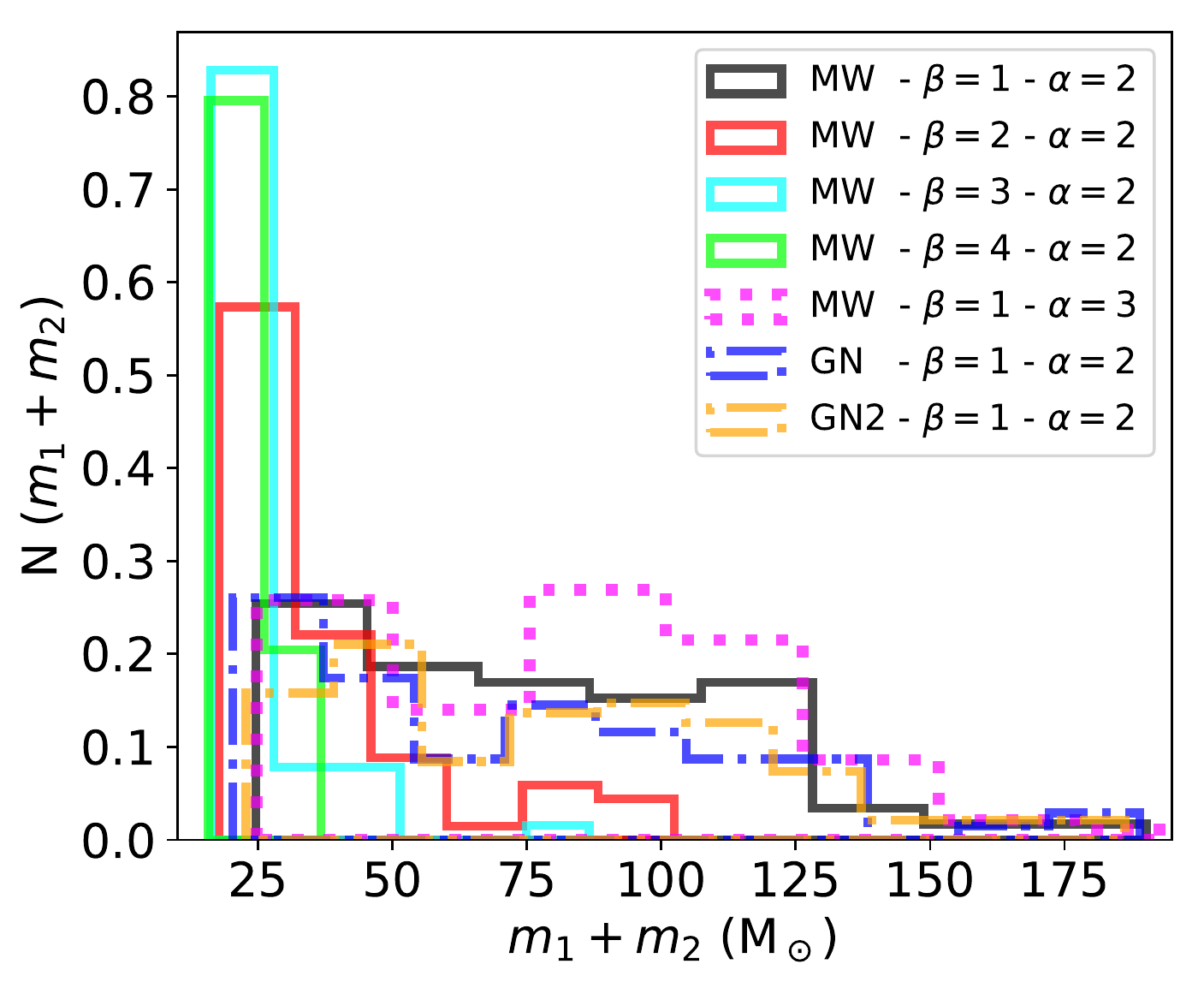}
\hspace{1cm}
\includegraphics[scale=0.55]{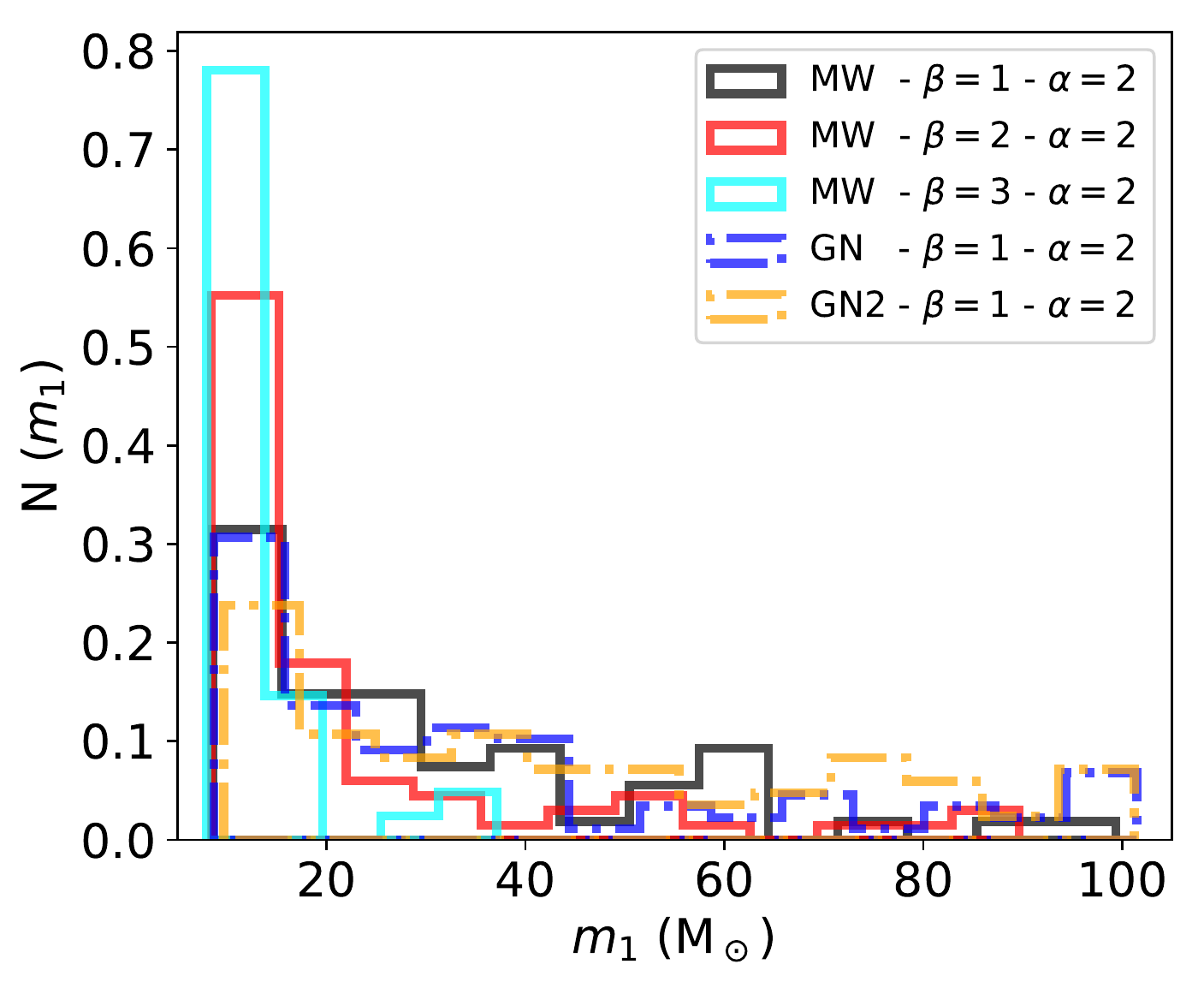}
\end{minipage}
\caption{Distribution of the total ($m_1+m_2$) BH-BH mass (left) and primary BH mass ($m_1$) in BH-NS binaries (right) for all models with $f(\ain)$ from \citet{hoan18}. As expected, the main parameter that affects the mass distribution is the slope $\beta$ of the BH mass function.}
\label{fig:masstot}
\end{figure*}

In this section, we use $N$-body simulations to study the fate of BH-BH, NS-NS and BH-NS binaries in galactic nuclei that host an SMBH. We consider three different SMBH masses, i.e. $\msmbh=4\times 10^6\msun$ for a Milky-Way-like nucleus (Models MW), $\msmbh=10^8\msun$ for a M31-like nucleus (Models GN) and $\msmbh=10^9\msun$ for more massive host galaxies (Models GN2). For the inner semi-major axis and eccentricities, we follow the prescriptions given by \citet{hoan18}, but also run some models with the sampling suggested by \citet{antoper12} to check how the initial conditions affect the final rates. Given the set of initial parameters as described in Sect.~\ref{sect:bhnsnuclei}, we draw the main parameters of the three-body system and require that the inner binary does not cross the Roche limit of the SMBH at its orbital pericentre distance
\begin{equation}
\frac{\aout}{\ain}> \eta \frac{1+\ein}{(1-\eout)}\left(\frac{3\msmbh}{m_1+m_2}\right)^{1/3}\,.
\label{eqn:hills}
\end{equation}
Following \citet{antoper12}, we set $\eta=4$ since at shorter distances the inner binary is unstable. We then integrate the triple SMBH-CO-CO differential equations of motion
\begin{equation}
{\ddot{\textbf{r}}}_i=-G\sum\limits_{j\ne i}\frac{m_j(\textbf{r}_i-\textbf{r}_j)}{\left|\textbf{r}_i-\textbf{r}_j\right|^3}\ ,
\end{equation}
with $i=1$,$2$,$3$. The integrations are performed using the \textsc{archain} code \citep{mik06,mik08}. This code is fully regularized and is able to model the evolution of objects of arbitrary mass ratios and eccentricities with extreme accuracy, even over long periods of time. We include PN corrections up to order PN2.5. 

For each set of parameters in Tab.~\ref{tab:models}, we run $\sim 1500$ simulations up to a maximum integration time of $T=1$ Myr, for a total of $\sim 35000$ simulations. From the computational point of view, this limit represents a good compromise between the numerical effort (large mass ratios and GW effects slow down the code) and the size of the statistical sample we want to take into account. From the physical point of view, we note that our total integration time is smaller than the typical timescale for vector resonant relaxation to operate \citep[$\sim$ few Myr, see Eq.~\ref{eqn:tvrr};][]{rauch96,kocs15}, which reorients the binary centre-of-mass orbital plane with respect to the SMBH, thus affecting the relative inclination of the inner and outer orbits and the relative LK dynamics, rendering the $3$-body approximation insufficient \citep{hamer18}. Furthermore, in-plane precession induced by the nuclear cluster potential and departure from spherical symmetry of the galactic nucleus would make the CO center of mass orbit precess even faster than vector resonant relaxation alone in a MW-like nucleus \citep{petr17}. Finally, we also note that our total integration time is smaller than the typical evaporation time of the CO binaries in galactic nuclei (see Eq.~\ref{eqn:binevap}). Taken all together, these considerations justify our choice of maximum integration time. Nevertheless, we also take into account in our rate calculations the binaries that are not affected by the LK cycles within $1$~Myr and merge by emission of GWs alone on longer timescales, without the assistance of LK oscillations\footnote{Note that these binaries could however be significantly perturbed by dynamical interactions, even before evaporating \citep{leigh16}}. Thus, our estimations correspond to a lower limit. We consider in total $23$ different models, which take into account different COs in the inner binary (BH-BH, BH-NS and NS-NS), different masses of the SMBHs, different slopes of the BH/NS mass functions, different spatial distributions of the CO binaries, and different inner semi-major axis and eccentricity distributions. Table~\ref{tab:models} summarizes all the models considered in this work.

In our simulations the CO binary has three possible fates: (i) the CO binary can survive on an orbit perturbed with respect to the initial one; (ii) the CO binary can be tidally broken apart by differential forces exerted by the SMBH, and its components will either be captured by the SMBH or ejected from the galactic nucleus; (iii) the CO binary merges producing an GW merger event. We distinguish among these possible outcomes by computing the mechanical energy of the CO binary. If the relative energy remains negative, we consider the binary survived (case (i)), otherwise we consider the binary unbound (case (ii)). Finally, if the CO binary merges, which occurs if the relative radii of the two COs overlap directly, we have a GW merger event (case (iii)).

\begin{table}
\caption{BH masses already detected via GW emission.}
\centering
\begin{tabular}{lcc}
\hline
Name &	$m_1$	(M$_\odot$) &	$m_2$	(M$_\odot$) \\
\hline\hline
GW150914  & $36.0^{+5.0}_{-4.0}$  & $29.0^{+4.0}_{-4.0}$ \\
GW151226  & $14.2^{+8.3}_{-3.7}$  & $7.5^{+2.3}_{-2.3}$ \\
GW170104  & $31.2^{+8.4}_{-6.0}$  & $19.4^{+5.3}_{-5.9}$ \\
GW170608  & $12.0^{+7.0}_{-2.0}$  & $7.0^{+2.0}_{-2.0}$ \\
GW170814  & $30.5^{+5.7}_{-3.0}$  & $25.3^{+2.8}_{-4.2}$ \\
LVT151012 & $23.0^{+18.0}_{-6.0}$ & $13.0^{+4.0}_{-5.0}$ \\
\hline
\end{tabular}
\label{tab:bhligo}
\end{table}

\subsection{Inclination distribution}

We illustrate in Fig.~\ref{fig:incl} the inclination as a function of semi-major axis $\ain$ and total mass in merged BH-BH binaries in our simulations for all the models with $f(\ain)$ from \citet{hoan18}. Most of the BH-BH binaries that merge have initial inclinations $\sim 90^\circ$, where the enhancement of the maximum eccentricity is expected to be larger due to LK oscillations. As discussed in Sect.~\ref{sect:secular}, the exact LK window angle depends on the physical quantities of the system \citep{grish17,grish18}, and the final distribution of surviving systems lacks highly inclined binaries \citep{fraglei18}. The lack of highly inclined systems shows the importance of the LK mechanism, since BH-BH binaries that successfully undergo a merger event originally orbit in a plane highly inclined with respect to the outer orbital plane \citep{fraglei18}. In these binaries, the LK mechanism influences the dynamics of the system and induces oscillations both in eccentricity and inclination, whenever not suppressed by GR precession. Figure~\ref{fig:incl} also shows that some systems that merge have inclinations far from $\sim 90^\circ$, in particular when the total binary mass is large and the inner semi-major axis is small. Also, these systems typically have relatively high initial inner eccentricities. 

\begin{figure*} 
\centering
\begin{minipage}{20.5cm}
\hspace{0.5cm}
\includegraphics[scale=0.50]{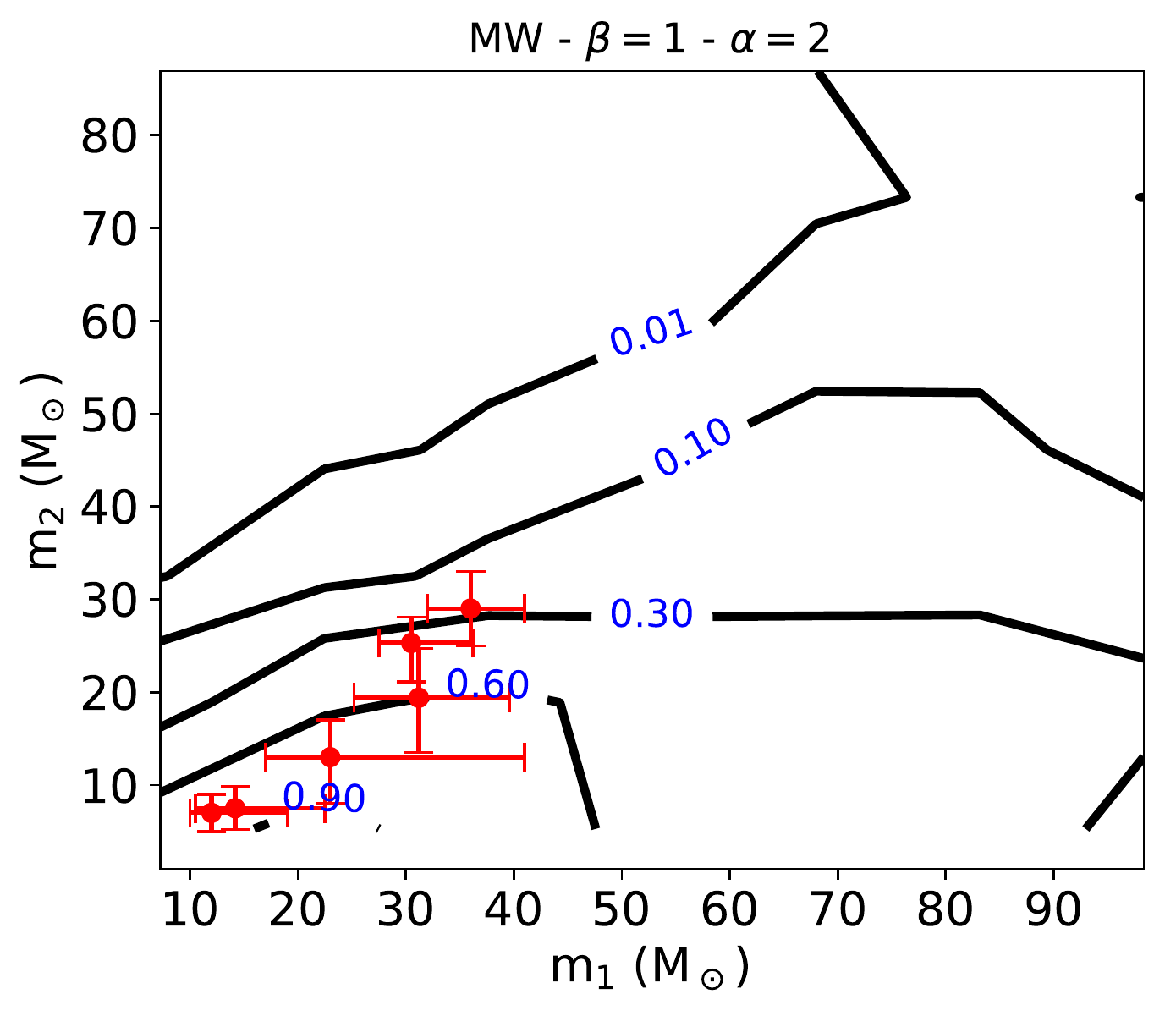}
\hspace{1.5cm}
\includegraphics[scale=0.50]{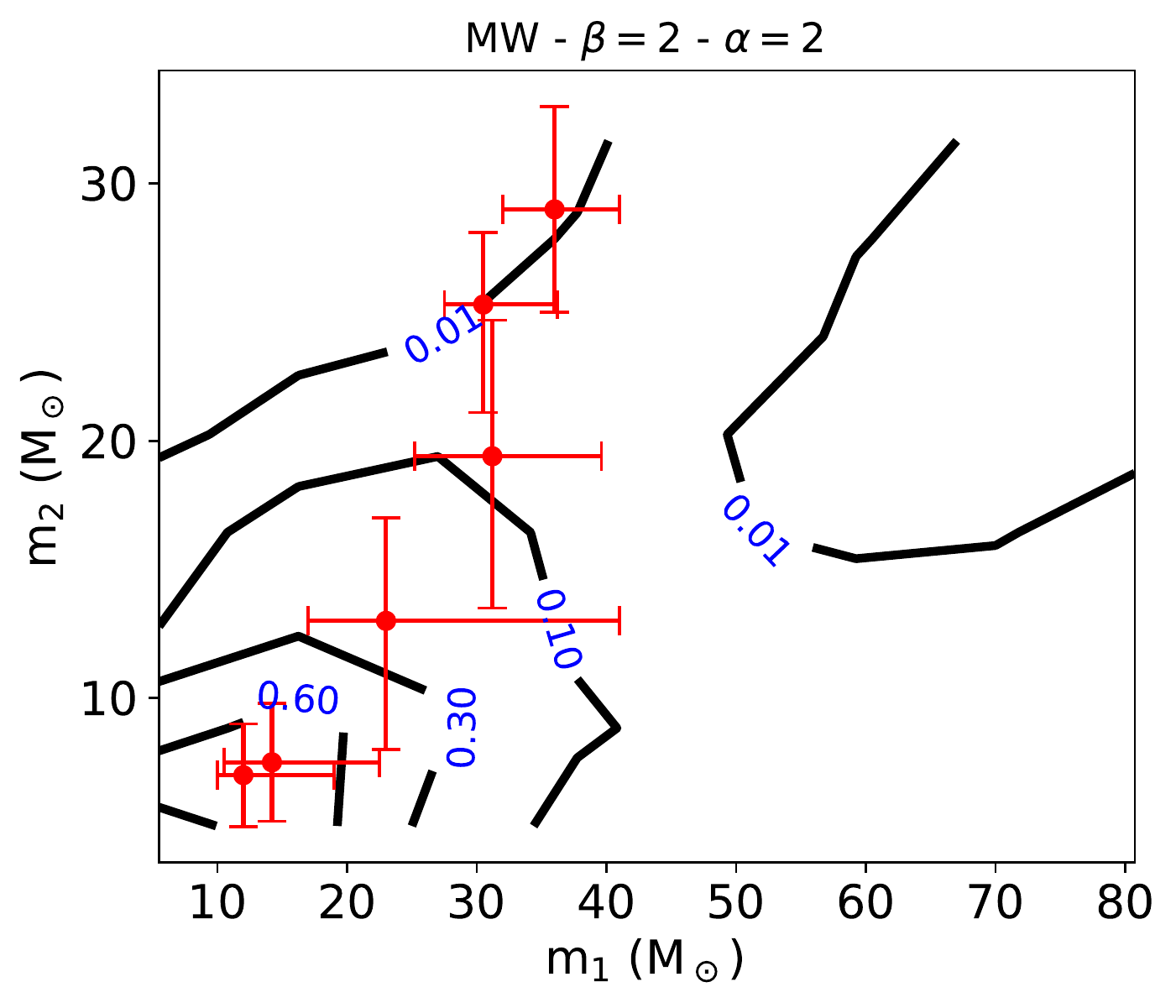}
\end{minipage}
\begin{minipage}{20.5cm}
\hspace{0.5cm}
\includegraphics[scale=0.50]{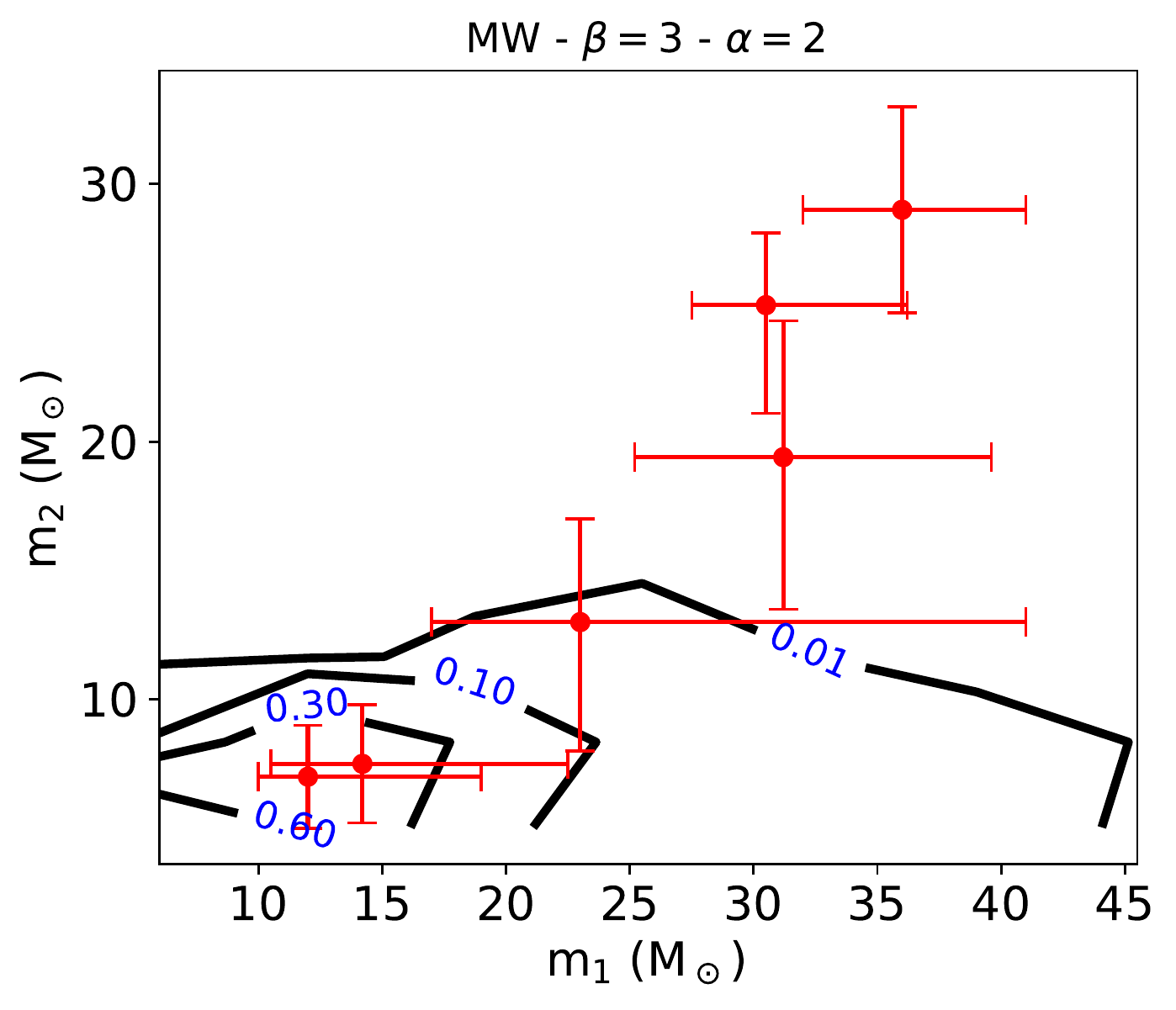}
\hspace{1.5cm}
\includegraphics[scale=0.50]{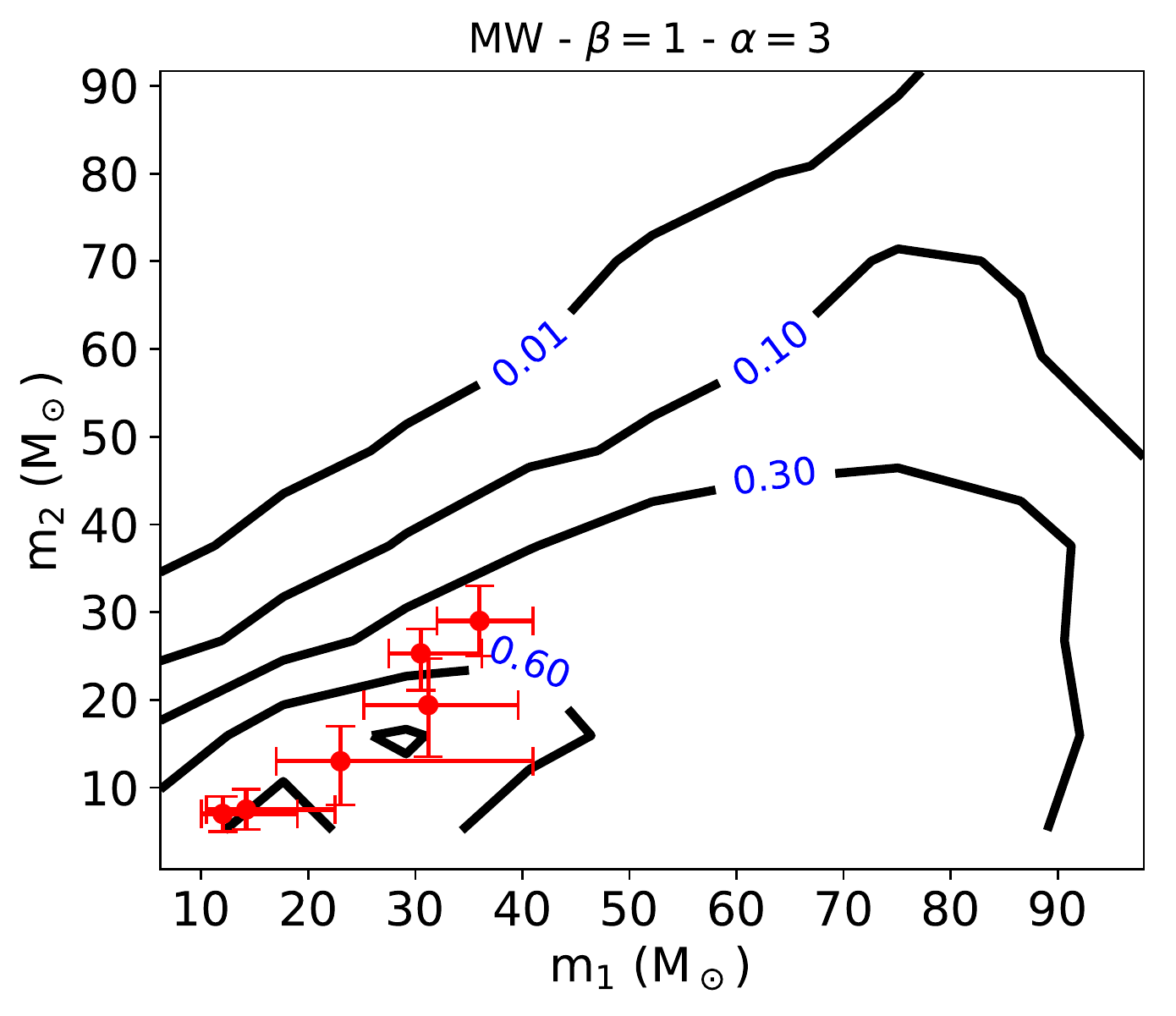}
\end{minipage}
\begin{minipage}{20.5cm}
\hspace{0.5cm}
\includegraphics[scale=0.50]{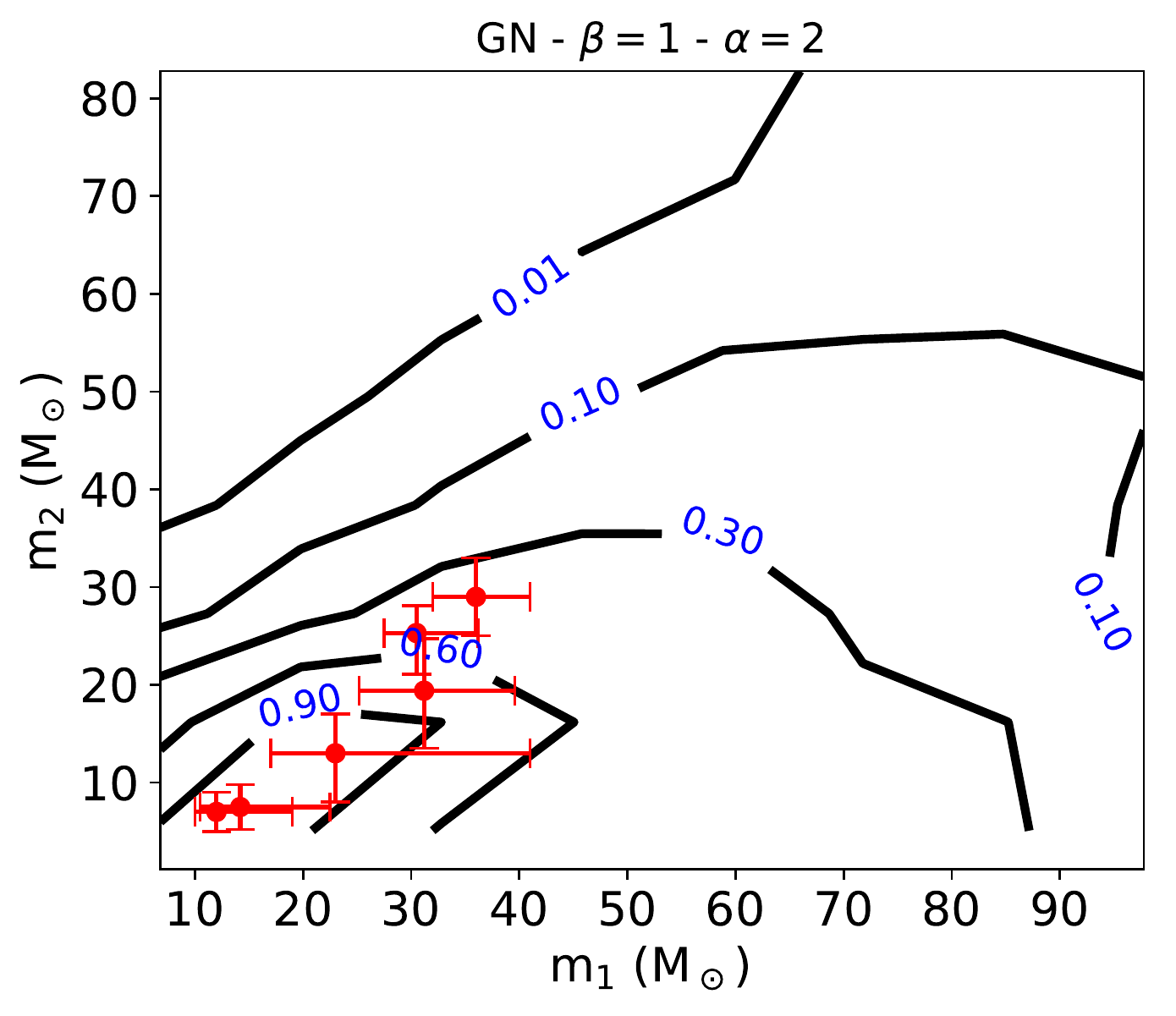}
\hspace{1.5cm}
\includegraphics[scale=0.50]{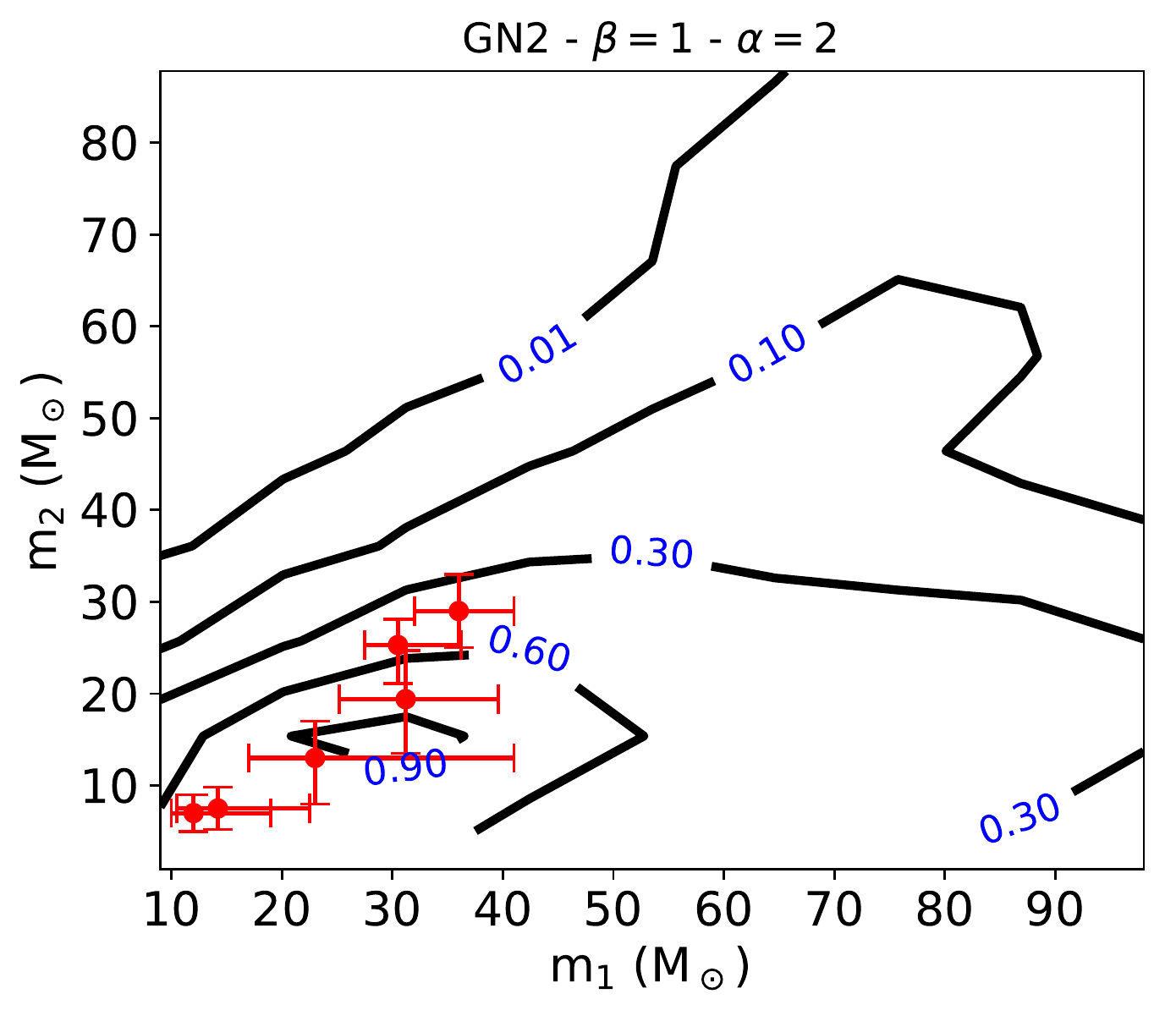}
\end{minipage}
\caption{Density maps of the masses of the BH-BH binaries that merge in our models (with $f(\ain)$ from \citet{hoan18}), along with a comparison to the observed BH masses detected via GW emission \citep[see Tab.~\ref{tab:bhligo};][]{abbott16a,abbott16b,abbott17,abbott17a,abbott17b}.}
\label{fig:massbhdata}
\end{figure*}

\subsection{Inner and outer orbital parameter distributions}

We present in Fig.~\ref{fig:ainaout} the cumulative distribution of $\ain$ (left panel) for BH-BH (top), BH-NS (centre), NS-NS (bottom) binaries for all models with $f(\ain)$ from \citet{hoan18}. Both the SMBH mass and the slope $\alpha$ of the CO binary spatial distribution around the SMBH significantly affect the inner and outer semi-major axes of merging binaries, as a consequence of their Hill stability \citep{grish17}. Larger SMBH masses imply that, on average, smaller values of $\ain$ are needed to avoid tidal disruption of the CO binaries after only a few orbits about the SMBH. For the same reason, large SMBH masses typically produce mergers at larger distances from the SMBH. Obviously, steep binary distributions (i.e., large $\alpha$'s) imply that the simulated CO binaries will, on average, be closer to the SMBH when they merge.  Hence, smaller inner semi-major axes are needed to avoid tidal dissociation by the SMBH. Finally, the total mass of the CO binary plays some role in shaping the final outer orbital semi-major axis distribution. CO binaries with smaller total masses ($m_1+m_2$) typically merge at larger distances from the SMBH, since their binding energy is easily overcome by the gravitational pull of the SMBH, which tends to break the binaries at smaller distances. For BHs, this translates into steeper mass functions typically producing mergers at farther distances from the SMBH.

\subsection{Mass distribution}

Figure~\ref{fig:masstot} shows the distribution of the total ($m_1+m_2$) BH-BH mass (left) and the primary BH mass ($m_1$) in BH-NS binaries (right) that merge in all models with $f(\ain)$ from \citet{hoan18}. The resulting mass distribution is barely affected by the slope of the binary spatial distribution around the SMBH, with a roughly constant shape in the range $\sim 25\msun- 125\msun$ and a tail extending up to $\sim 180\msun$ for BH-BH binaries.  Also, the mass of the central SMBH does not significantly affect the mass distribution. As expected, the parameter that governs the resulting shape of the mass distribution is the slope $\beta$ of the BH mass function: the shallower the BH mass function, the larger the typical total mass of merging BH-BH binaries. In the case $\beta=1$, we find that $\sim 95\%$ of the mergers have $m_1+m_2\lesssim 150\msun$, while roughly all the mergers have total masses $\lesssim 100\msun$, $\lesssim 50\msun$, and $\lesssim 25\msun$ for $\beta=2$, $\beta=3$, and $\beta=4$, respectively. Similar results also hold for BH-NS binaries.

The slope of the BH mass function is unknown. We can use the results of our simulations along with the BH-BH merger events observed by LIGO \citep[see Tab.~\ref{tab:bhligo};][]{abbott16a,abbott16b,abbott17,abbott17a,abbott17b} to constrain the BH mass function, assuming these mergers took place in a galactic nucleus. We show in Fig.~\ref{fig:massbhdata} density maps for the masses of the two merging BHs ($m_1>m_2$), along with data from the LIGO-observed BH merger events. It is clear that a steep mass function ($\beta>1$) seems to be disfavored by the current data, which suggest a shallower BH mass function. However, we note that BH mass measurements via GW  observations are  biased towards more massive BHs, since these are more easily observed by LIGO. We also note that, although the mass distribution is only slightly affected by the SMBH mass, the data seem to prefer more massive nuclei than the Milky-Way. Note, however, that mass-segregation processes, that can give rise to much steeper effective mass-functions of BHs in galactic nuclei \citep{aharon16}, only operate in small-SMBH nuclei where relaxation (and mass-segregation) times are short.

\subsection{Eccentricity}

\begin{figure}
\begin{centering}
\includegraphics[scale=0.5]{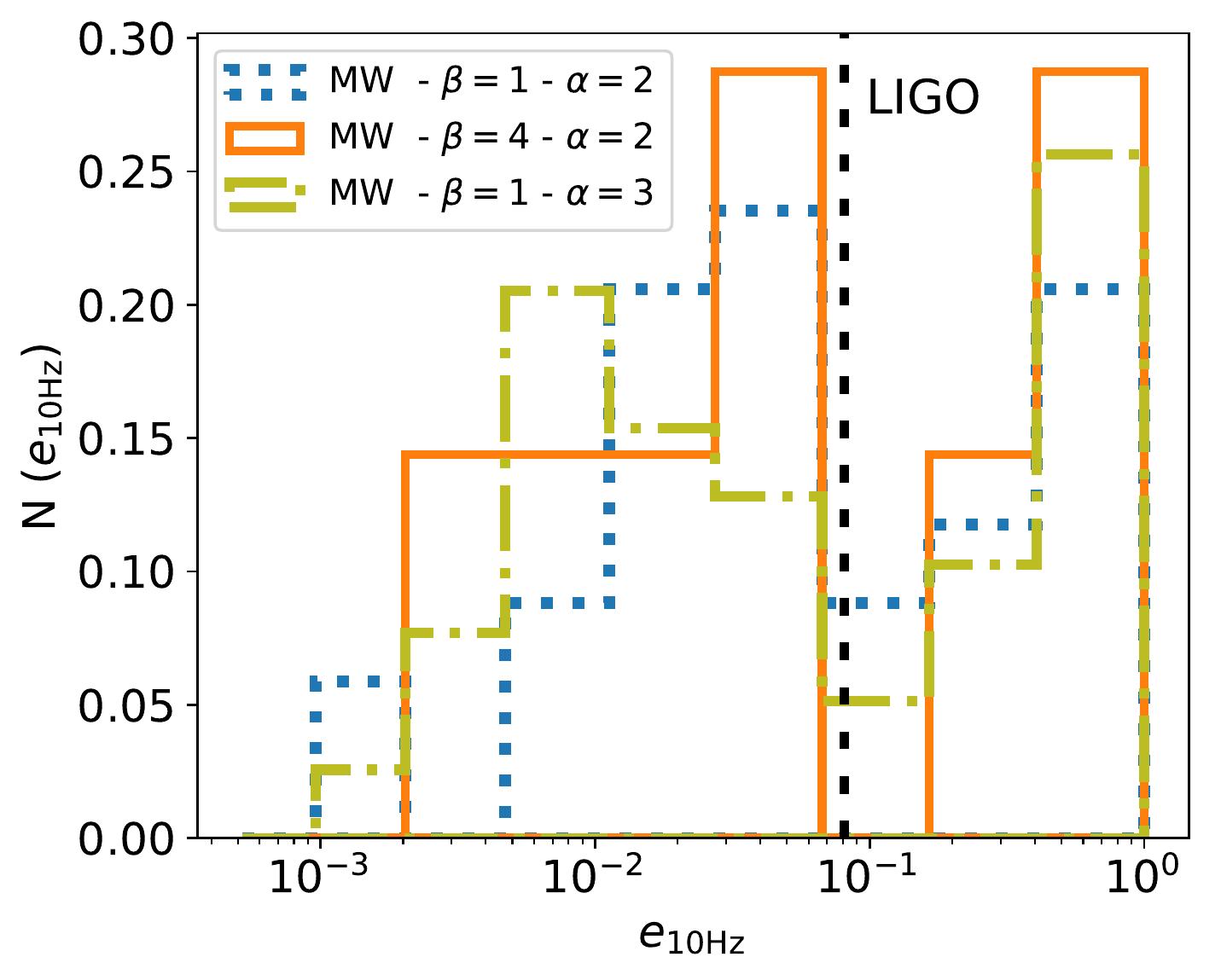}
\par\end{centering}
\caption{Distribution of eccentricities at the moment the binaries enter the LIGO frequency band ($10$ Hz) for BH-BH mergers in a Milky Way-like nucleus, for different values of $\beta$ and $\alpha$. The vertical line shows the minimum $e_{\rm 10Hz}=0.081$ where LIGO/VIRGO/KAGRA network may distinguish eccentric sources from circular sources \citep{gond2019}. A significant fraction of binaries have a significant eccentricity in the LIGO band.}
\label{fig:eccligo}
\end{figure}

For the systems that merge in our simulations, we compute a proxy for the GW frequency of the merging binaries.  This is taken to be the frequency corresponding to the harmonic that gives the maximal emission of GWs \citep{wen03}
\begin{equation} 
f_{\rm GW}=\frac{\sqrt{G(m_1+m_2)}}{\pi}\frac{(1+e_{\rm in})^{1.1954}}{[\ain(1-e_{\rm in}^2)]^{1.5}}\ .
\end{equation}
Figure~\ref{fig:eccligo} reports the distribution of eccentricities at the moment the binaries enter the LIGO frequency band ($10$ Hz) for BH-BH mergers in a Milky Way-like nucleus, for different values of $\beta$ and $\alpha$. The distributions have a double peak at $e_{\rm 10Hz}\sim 10^{-2}$ and $e_{\rm 10Hz}\sim 1$. In our model where we take $a_{\rm out}^{M}=0.5$ pc, we find a similar distribution of eccentricities. Binaries merging in galactic nuclei typically have larger eccentricities than those formed through most other channels, particularly mergers in isolated binary evolution and in SBHBs ejected from star clusters. However, mergers that follow from the GW capture scenario in clusters \citep{zevin18} and galactic nuclei \citep{gondan2018,rass2019}, from resonant binary-single scattering in clusters \citep{sam18}, from hierarchical triples \citep{ant17,fragk2019,flp2019,fralo2019}, and from BH binaries orbiting intermediate-mass black holes in star clusters \citep{fragbr2019} also present a similar peak at high eccentricities. We find that typically $\sim 20$--$30\%$ of binaries have $e\gtrsim 0.1$ in the LIGO band \citep{gond2019}. In our runs, we also find that some of the CO binaries that merge do not merge due to LK oscillations, but instead merge by emission of GWs on timescales longer than $1$ Myr and with eccentricities at $10$ Hz much smaller than the typical eccentricity reported in Figure~\ref{fig:eccligo}. Their relative fraction is typically $\sim 20\%$-$50\%$ of the total mergers we find in our simulations. When their contribution is taken into account, the fraction of binaries that enter the LIGO band with $e\gtrsim 0.1$ decreases to $\sim 10\%$-$20\%$. This fraction is still larger than previously estimated values ($\sim 1$\%) found in the literature \citep{antoper12,van16,rand18}, probably due to the different integration schemes adopted. In a secular approach, the averaged equations of motion could smear out the peak at high eccentricities and lower the number of binaries entering the LIGO band with very high eccentricities.

Finally, the high eccentricities we find in those binaries that merge may imply that a fraction of these binaries could emit their maximum power at higher frequencies, possibly in the range of LISA.

\section{Merger time distributions and rates}
\label{sect:timedistrates}

\begin{figure*} 
\centering
\begin{minipage}{20.5cm}
\hspace{0.5cm}
\includegraphics[scale=0.5]{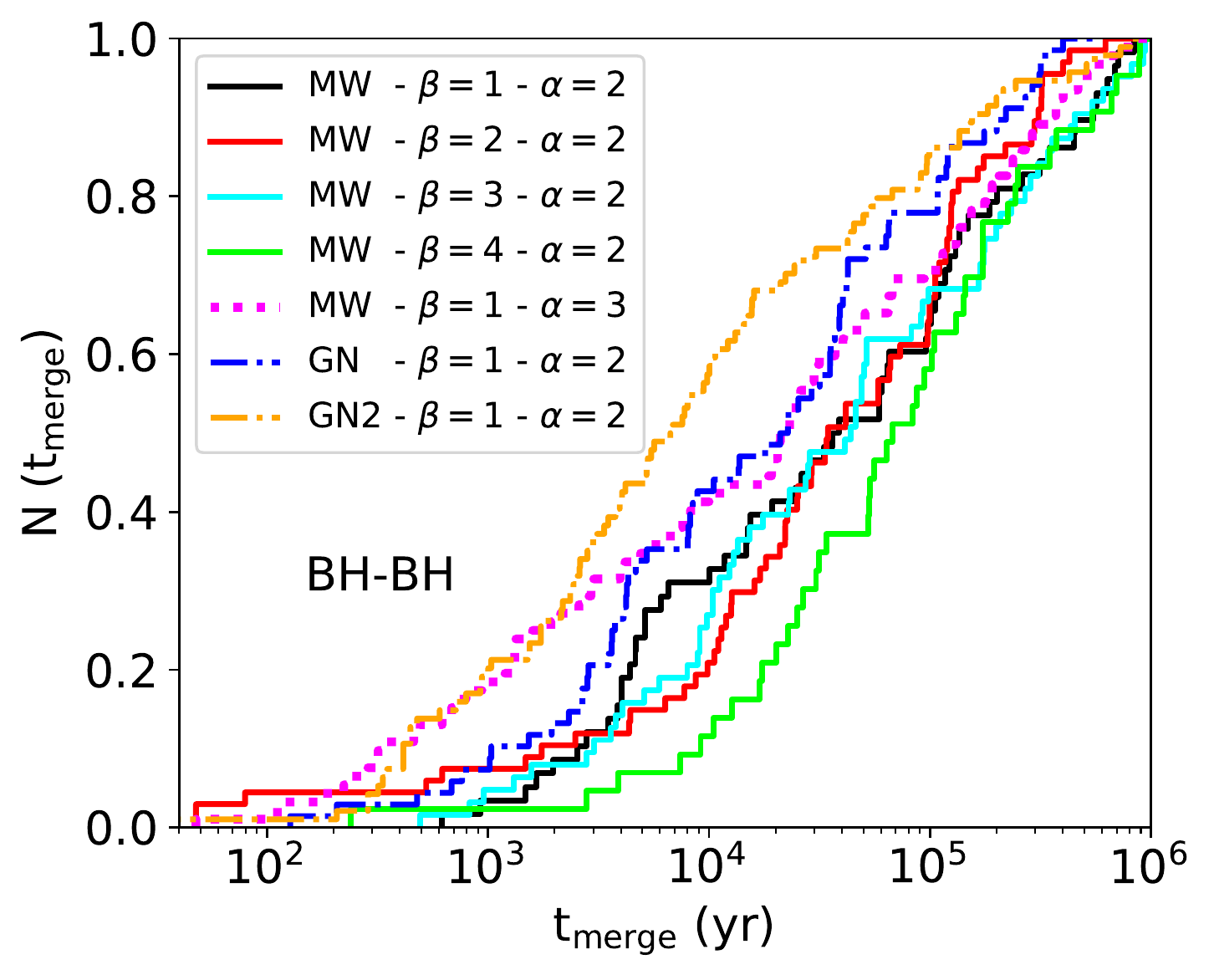}
\hspace{1.5cm}
\includegraphics[scale=0.5]{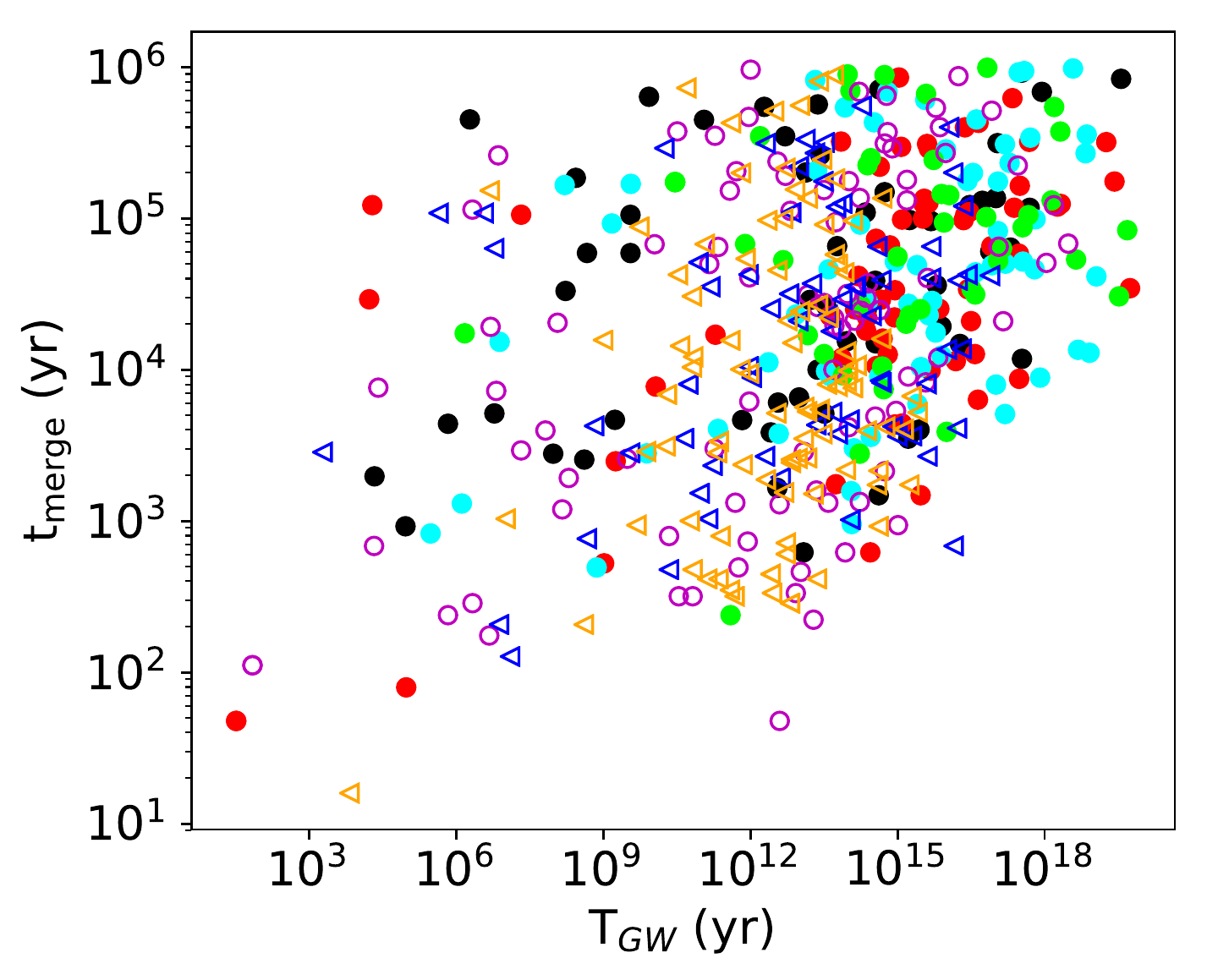}
\end{minipage}
\begin{minipage}{20.5cm}
\hspace{0.5cm}
\includegraphics[scale=0.5]{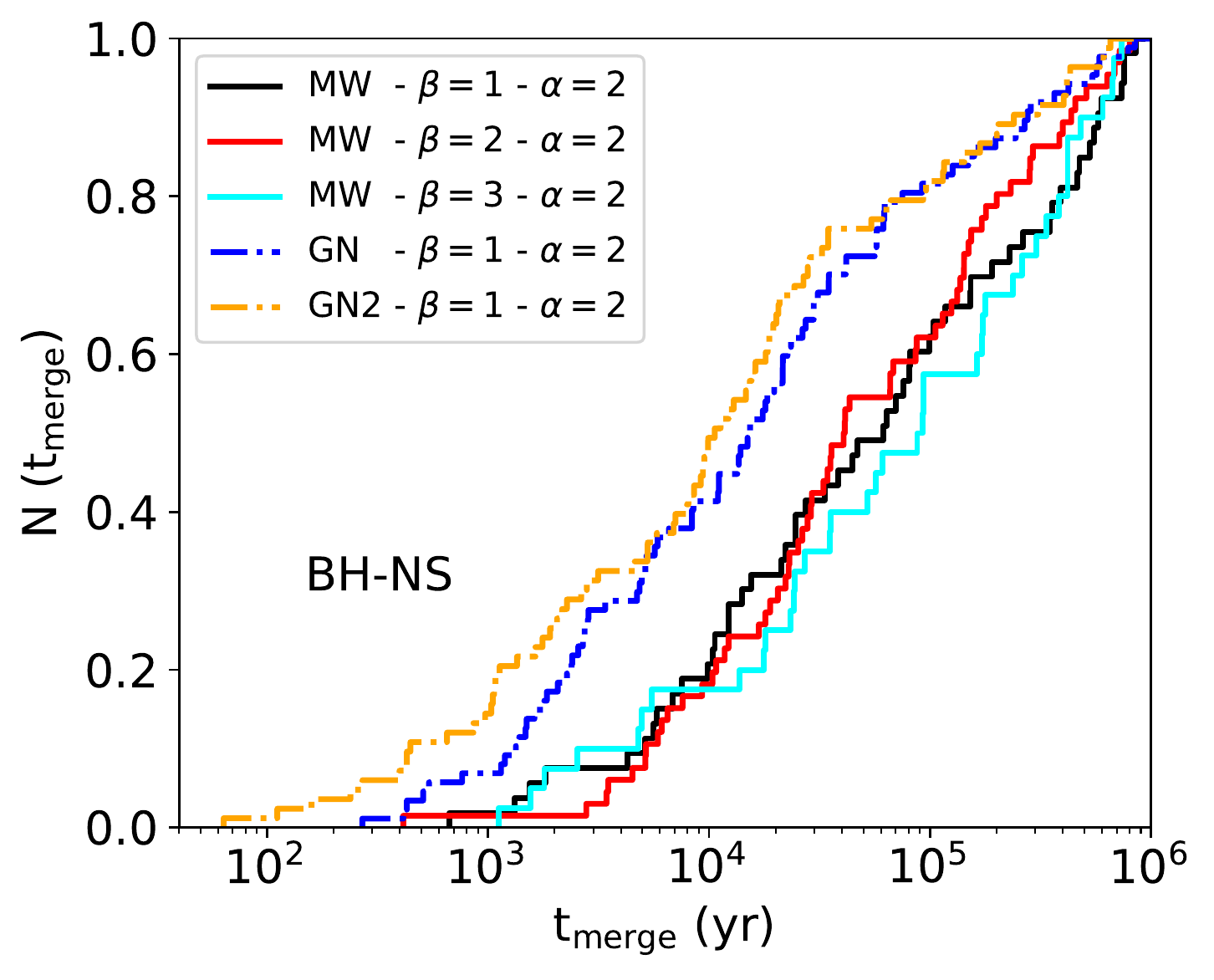}
\hspace{1.5cm}
\includegraphics[scale=0.5]{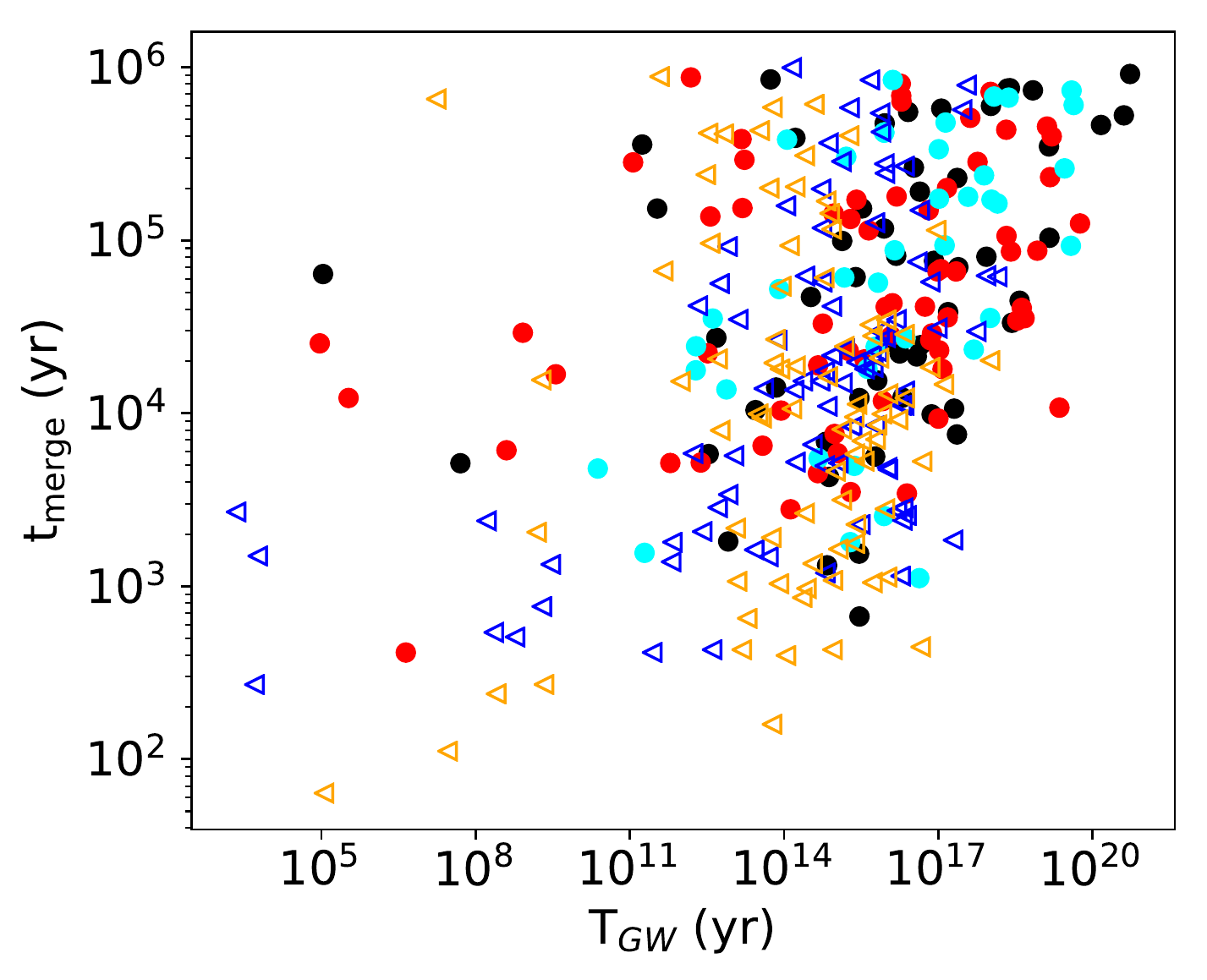}
\end{minipage}
\begin{minipage}{20.5cm}
\hspace{0.5cm}
\includegraphics[scale=0.5]{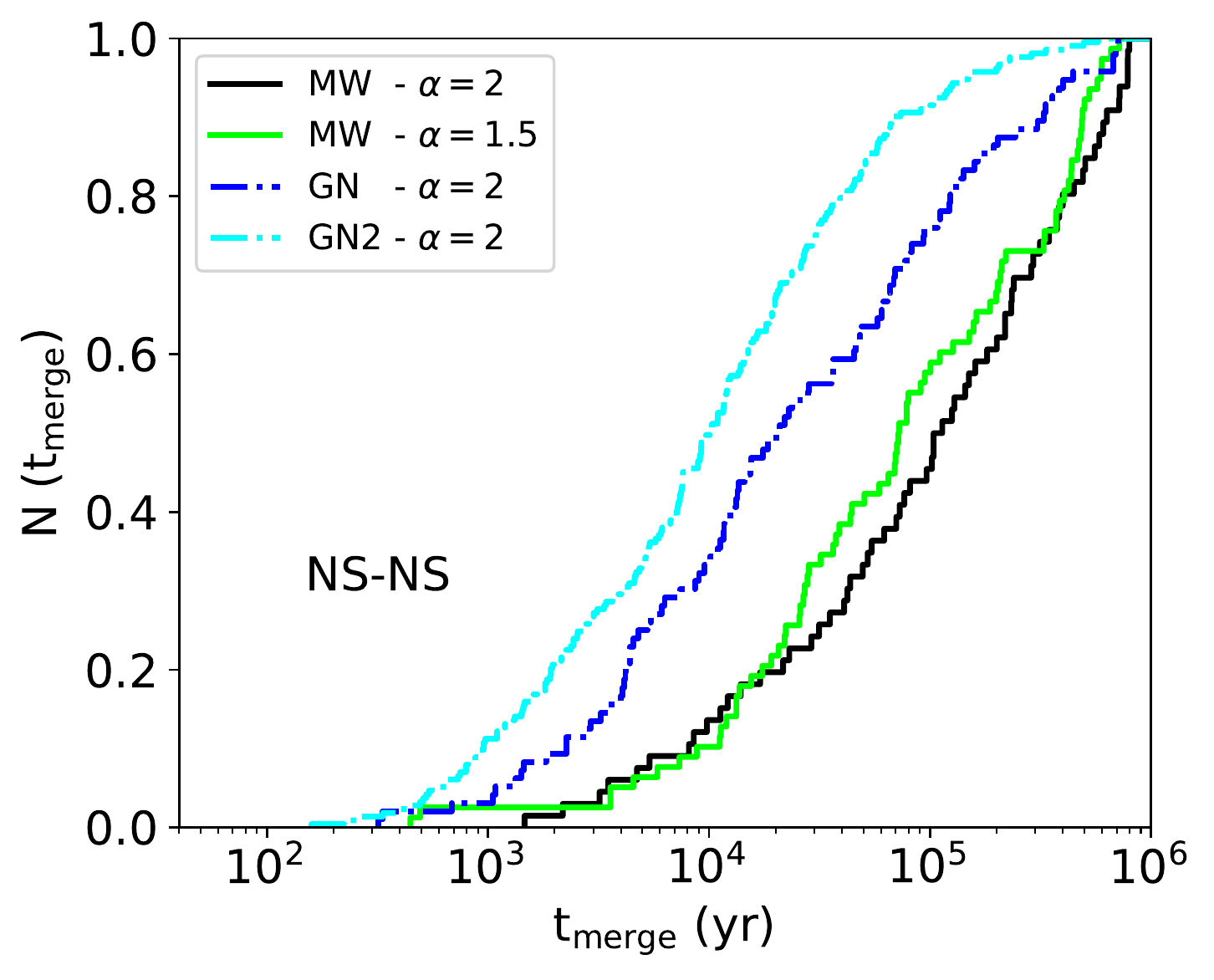}
\hspace{1.5cm}
\includegraphics[scale=0.5]{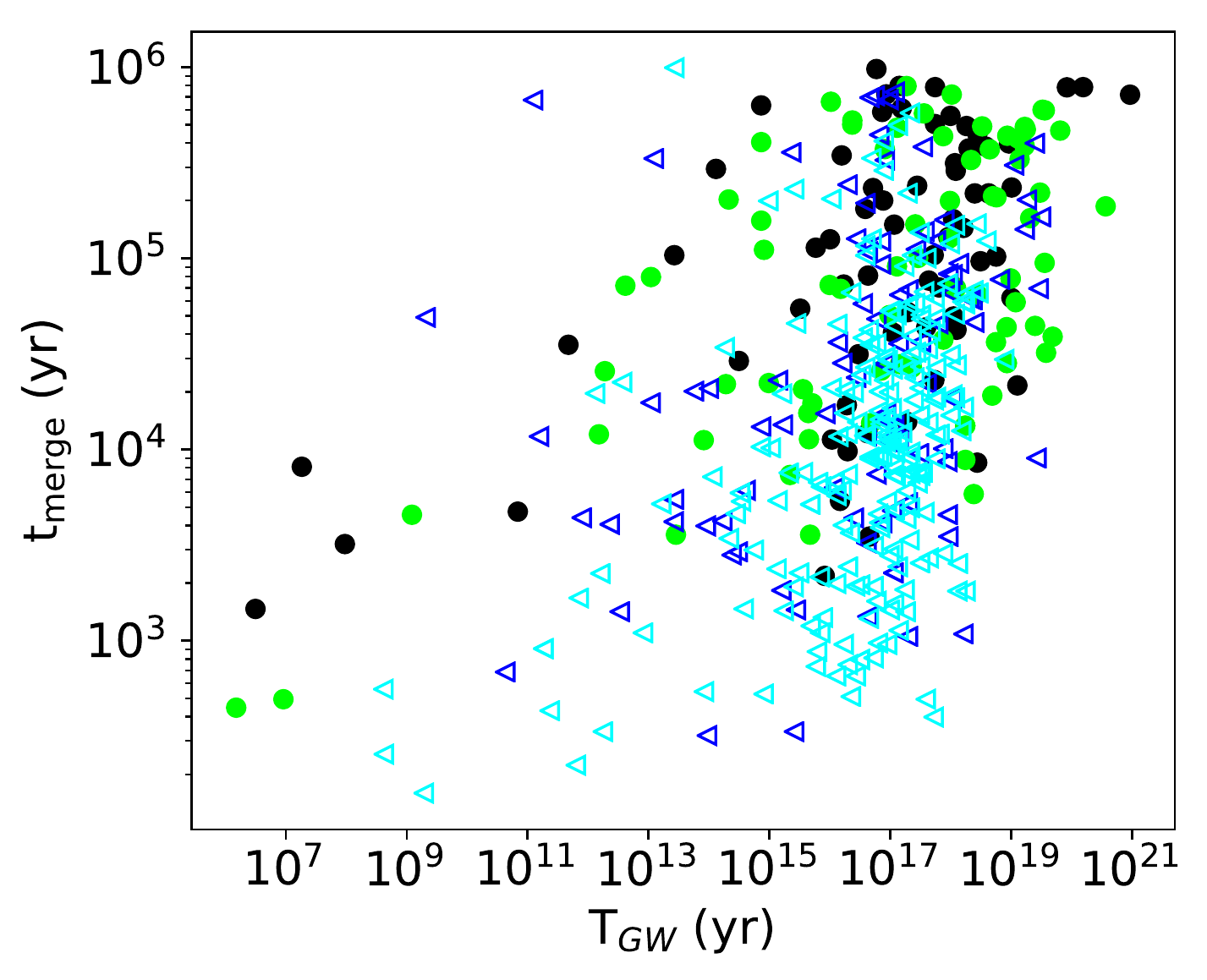}
\end{minipage}
\caption{Left panel: cumulative merger time ($\tmerg$) distribution for BH-BH (top), BH-NS (centre) and NS-NS (bottom) binaries, for all models with $f(\ain)$ from \citet{hoan18}. Right panel: $\tmerg$ as a function of the nominal \citet{pet64} GW merger time-scale $T_{GW}$, for the same models as shown in the left panel.}
\label{fig:tmerg}
\end{figure*}

\begin{figure}
\includegraphics[width=8cm]{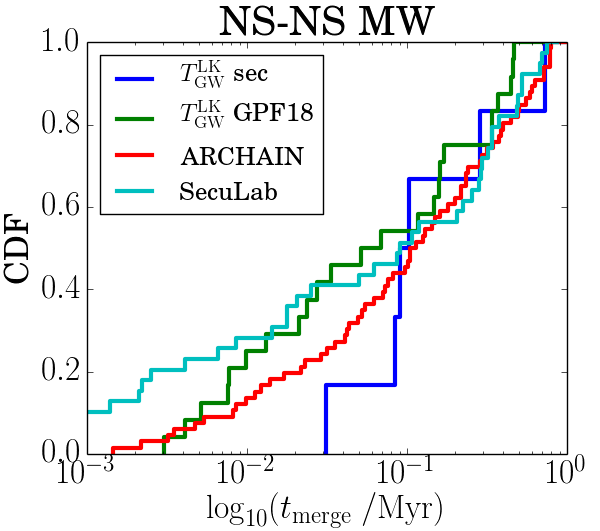}
\caption{\label{fig:tdist} Cumulative distribution of merger times for merged orbits. Red: \textsc{archain} N-body simulation. The analytic merger estimate for the \textsc{archain} initial conditions is from Eq. \ref{eq:T_LKGW} is shown in blue (secular approximation) and green (GPF18 correction, see text).  Cyan: \textsc{SecuLab} secular simulation. The distribution of initial conditions is the same but the sample is larger ($N=10^4$).}
\end{figure}

As discussed, the SMBH plays a fundamental role in reducing the merger timescale from the nominal value in Eq.~(\ref{eq:t_merge}) \citep{antoper12,hoan18,fraglei18b}. In Fig.~\ref{fig:tmerg}, we present for all models the cumulative distribution of merger times ($\tmerg$; left panel) for BH-BH (top), BH-NS (centre) and NS-NS (bottom) binaries. The merger time distribution is nearly independent of our assumptions for the BH mass function slope $\beta$ and the CO binary spatial distribution slope $\alpha$.  It depends only on the SMBH mass. Larger SMBH masses imply shorter merger times due to more intense perturbations from the SMBH. In the right panel of Fig.~\ref{fig:tmerg}, we show $\tmerg$ as a function of the nominal \citep{pet64} GW merger time-scale $T_{\rm GW}$, for all binaries that merge in our simulations. Due to oscillations in the orbital elements, the CO binaries merge much faster than predicted by Eq.~\ref{eq:t_merge}, by several orders of magnitude.

\subsection{Comparing secular techniques and N-body simulations}
\label{subsect:comparison}

In order to compare the distribution of merger times with different prescriptions, we take the initial conditions and calculate the merger time from Eq. (\ref{eq:T_LKGW}). We find the maximal eccentricity by solving Eq.~(\ref{eq:jmin_e0}) \citep{grish18}. For the secular case, we use $\epsilon_{\rm SA}=0$, which also implies $\delta e=0$ from Eq. (\ref{eq:emax_corr}), while for the corrected case, $\epsilon_{\rm SA} > 0 $ is set from the system's initial conditions. If the merger time is shorter than the secular LK timescale that is required to reach the maximal eccentricity, we use the secular LK time. We then filter out the merger times longer than $1\ \rm Myr$ and compare to the simulated merger times. 

We use the initial conditions of the MW NS-NS case with $\alpha=2$ to calculate $T_{\rm GW}^{\rm LK}$ as prescribed above. In addition, in order to compare with \textsc{archain}, we draw $N=10^4$ initial conditions from the same distribution and use a secular code to evolve them up to $T=1\ \rm Myr$ and record their merger times. For the secular code, we use \textsc{SecuLab}\footnote{https://github.com/eugeneg88/SecuLab}, a publicly available code that solves the secular equations of motion up to octupole order with additional secular 2.5PN terms. 

Table \ref{tab:tmerge_nsns} shows the expected merger fractions from the direct N-body and the secular codes, respectively, together with different methods of evaluating the merger time. $T_{{\rm GW}}^{{\rm LK}}$ is evaluated once using the secular approximation (sec), and the corrected eccentricity (\citealp{grish18}; GPF18). The merger rate is given by the fraction of initial conditions that result in $T_{{\rm GW}}^{{\rm LK}}\le 1\ \rm Myr$. We see that $T_{{\rm GW}}^{{\rm LK}}$ slightly overestimates the merger rate in the secular regime, but underpredicts it in the N-body regime.

Overall, the number of events from both simulations exceeds their expected analytic estimate. For the secular case, the ratio between the merger rate obtained numerically from \textsc{SecuLab} and analytically from Eq. (\ref{eq:T_LKGW}) is $0.004/0.0028=1.43$. Similarly, for the N-body case, this ratio is $0.028/0.01=2.8$. The ratio of merger rates for both simulations is about $0.028/0.004 = 7$. The possible origins of these discrepancies are discussed below.

\begin{table}
\caption{Fraction of merger for the MW NS-NS case. First and second row are the fractions from \textsc{SecuLab} and \textsc{archain}, respectively. Third and forth row are the expected mergers from evaluating Eq. \ref{eq:T_LKGW} with different maximal eccentricity evaluations (see text).}
\begin{center}
\begin{tabular}{|c|c|c|c|c|}
\hline 
 & SecuLab & ARCHAIN & $T_{{\rm GW}}^{{\rm LK}}$ sec & $T_{{\rm GW}}^{{\rm LK}}$ GPF18 \tabularnewline
\hline 
\hline 
$f_{{\rm merge}}$ & $0.004$ & $0.028$ & $0.0028$ & $0.01$\tabularnewline
\hline 
\end{tabular}
\par\end{center}
\label{tab:tmerge_nsns}
\end{table}

\begin{table}
\caption{$D$ and $p$ values for double-sided Kolmogorov-Smirnov (KS) tests of the cumulative distributions of the merger times.}
\begin{center}
\begin{tabular}{|c|c|c|}
\hline 
$(D,p)$ values & SecuLab & ARCHAIN\tabularnewline
\hline 
\hline 
$T_{{\rm GW}}^{{\rm LK}}$ sec & $0.4,0.23$ & $0.29,0.58$\tabularnewline
\hline 
$T_{{\rm GW}}^{{\rm LK}}$ GPF18 & $0.2,0.52$ & $0.19,0.5$\tabularnewline
\hline 
\end{tabular}
\par\end{center}
\label{tab:kstests}
\end{table}

Figure \ref{fig:tdist} shows the cumulative distribution of the merger times for the MW NS-NS case, where both secular and direct N-body simulations were used. Overall, the total number of mergers predicted by the corrected GPF18 model is larger, since larger eccentricities are involved, which is compatible with the numerical results. The secular model is less accurate with decreasing merger timescales; this is because the eccentricities involved are extreme, therefore deviations from the secular regime are more severe.

On longer merger timescales, $t_{\rm merge} \gtrsim 10^4 \ \rm yr$, if the typical LK timescales are short enough the maximal eccentricity attained, $e_{\rm max}$, is usually larger than the predicted one from 1PN theory alone (see their Fig. 5 of \citealp{grish18}). Therefore, the mergers occur of faster than expected and the CDF is underestimated. On the other hand, for short merger times, the maximal eccentricity of the GPF18 model unbound, and the merger occurs on the secular timescale. If, however, $t_{\rm LK} \sqrt{1 - e_{\rm max}^2} \lesssim P_{\rm in}$, the GPF18 model also breaks down and cannot describe the system \citep{ant17}, while the value of $e_{\rm max}$ is stochastic. This is because the fraction of time the inner orbit spends near $e_{\rm max}$ (i.e. $\sim t_{\rm LK} \sqrt{1 - e_{\rm max}^2}$) is too short for the inner orbit to complete one revolution and reach pericentre. Thus it takes longer time (at least a few secular times) to merge and the CDF is overestimated.

In order to get a better sense of the corrected prescription, we preform two sided Kolmogorov-Smirnov (KS) tests comparing the cumulative distribution function (CDF) from the simulated results versus the secular merger time distributions. Table \ref{tab:kstests} shows the resulting D values and p values upon comparing the $N$-body and secular simulations with the secular and the corrected GPF18 model. 

For the KS statistics, the D values are better for the GPF18 model, since the distance between the simulated and GPF18 CDFs is smaller. The p values are comparable for all models. A possible statistical artifact could be the low number of predicted mergers for the secular case. This suggesting that neither of the models is comparable with the simulated distributions.

To summarize, the simulated distribution cannot be fully described by the GPF18 model in a statistical sense, but the overall trend of larger merger fractions on shorter merger timescales is consistent with the simulations. Thus, further improvements in the analytic understanding both in the GPF18 model and in the \cite{pet64} formulae are highly desired and deserve future work.

\subsection{Merger Rates}
\label{subsect:rates}

\begin{table*}
\caption{Rates of CO binary mergers (in $\ \mathrm{Gpc}^{-3}\ \mathrm{yr}^{-1}$) as a function of the SMBH mass.}
\centering
\begin{tabular}{lcccc}
\hline
$\msmbh$ ($\msun$) & $\Gamma_{\rm BH-BH}$ & $\Gamma_{\rm BH-NS}$ & $\Gamma_{\rm NS-NS}$ & $\Gamma^{\rm SF}_{\rm NS-NS}$\\
\hline\hline
$4\times 10^6$  & $0.52$ & $0.10$ & $1.71\times 10^{-3}$ & $0.16$ \\
$10^8$ 			& $0.24$ & $0.08$ & $1.27\times 10^{-3}$ & $0.11$ \\
$10^9$ 			& $0.17$ & $0.06$ & $0.41\times 10^{-3}$ & $0.04$ \\
\hline
\end{tabular}
\label{tab:rates}
\end{table*}

Although we explore only a limited number of SMBH masses, we note that the range is relatively representative of what is expected for galactic nuclei. With the results of our simulations in hand, we can derive the expected merger rates of BH-BH, BH-NS and NS-NS binaries. With this, we can infer the dependence of the rate on the distribution of SMBH masses in the nearby Universe.  

Following \citet{hamer18}, we calculate the merger rate for CO binaries as
\begin{equation}
\Gamma(\msmbh)=n_{\rm gal} f_{\rm SMBH} \Gamma^{\rm sup}_{\rm CO} f_{\rm bin} f_{\rm merge}\ ,
\end{equation}
where $n_{\rm gal}$ is the galaxy density, $f_{\rm SMBH}\approx 0.5$ is the fraction of galaxies containing an SMBH \citep{anto15a,anto15b}, $\Gamma^{\rm sup}_{\rm CO}$ is the compact object supply rate, $f_{\rm bin}$ is the fraction
of stars forming compact object binaries, and $f_{\rm merge}$ is the fraction of mergers we find in our simulations. For galaxies, we assume that the SMBH number density scales as $\Phi(\msmbh)\propto 1/\msmbh$ \citep{aller02}, hence the integrated number density of galaxies scales as
\begin{equation}
n_{\rm gal}\propto \int \Phi(\msmbh) d\msmbh \propto \log(\msmbh)\ .
\end{equation}
As in \citet{hamer18}, we neglect the weak dependence on the SMBH mass and fix $n_{\rm gal}=0.02$ Mpc$^{-3}$ \citep*{cons05}. 

The fraction of CO binaries in the GC strongly depends on the assumptions regarding their origins. Several possibilities have been discussed in \citet{antoper12}; here we focus on two: ex-situ and in-situ origins. In the ex-situ scenario, stars form outside the nuclear cluster and then diffuse inwards. In the in-situ formation scenario, stars are formed in-situ close to the SMBH. We use simplified assumptions to estimate the supply rate in both cases. For a relaxed nuclear cluster, Fokker-Planck, Monte-Carlo and N-body simulations suggest that the fractions of BHs and NSs in the central 0.1 pc are of the order of $\gamma_{\rm CO}=0.06,\,0.01$ for BHs and NSs, respectively (the higher BH fractions are due to mass-segregation), assuming the background stellar population has a continuous star-formation rate \citep{hopm06}. Following \cite{antoper12} we take initial binary fractions of $f_{\rm bin}=0.1,\,0.07$ for BHs and NSs, respectively. For the compact object formation rate, we assume that the compact objects are supplied to the galactic nucleus by $2$-body relaxation and mass segregation
\begin{equation}
\Gamma^{\rm sup}_{\rm CO}=\frac{\gamma_{\rm CO} N_* (0.1\ \mathrm{pc})}{t_{\rm seg}(0.1\ \mathrm{pc})}\ \propto \msmbh^{(3-\beta)/\beta}\ ,
\end{equation}
where $\gamma_{\rm CO}$ is the fractional number of compact objects; $t_{\rm seg} = T_{\rm 2b}(m_{\rm bin}/{\rm M_\odot})$ is the timescale for mass segregation for binaries with mass $m_{\rm bin}$; and we assume $\msmbh\propto \sigma^4$ \citep{merr01}.

Normalizing the rates to the Milky Way's Galactic Centre
\begin{equation}
\Gamma^{\rm sup}_{\rm BH}=2.5\times10^{-6} \left(\frac{4\times 10^6\msun}{\msmbh} \right)^{1/4}  \mathrm{yr}^{-1}\ .
\end{equation}
\begin{equation}
\Gamma^{\rm sup}_{\rm NS}=2.3\times10^{-8} \left(\frac{4\times 10^6\msun}{\msmbh} \right)^{1/4}  \mathrm{yr}^{-1}\ .
\end{equation}
The final expression for our rate becomes
\begin{equation}
\Gamma_{\rm BH}(\msmbh)= 3.5f_{\rm merge}\ \mathrm{Gpc}^{-3}\ \mathrm{yr}^{-1} \times\left(\frac{4\times 10^6\msun}{\msmbh} \right)^{1/4} ,
\label{eqn:ratebh}
\end{equation}
\begin{equation}
\Gamma_{\rm NS}(\msmbh)= 3.2\times 10^{-2} f_{\rm merge}\ \mathrm{Gpc}^{-3}\ \mathrm{yr}^{-1} \times\left(\frac{4\times 10^6\msun}{\msmbh} \right)^{1/4} ,
\end{equation}
which is weakly dependent on the SMBH mass, and where the merger fraction $f_{\rm merge}$, which is typically a few up to a few tens of percents for the various models we considered, can be found in Tab.~\ref{tab:models}. We note that in $f_{\rm merge}$ we have also included the CO binaries that would merge by emission of GWs in timescales $>1$ Myr, without the assistance of LK oscillations. These are typically a few percent of the total mergers in the case of BH-NS and NS-NS binaries, while $\sim 20$-$50\%$ in the case of BH-BH binaries.

Note that the evaporation time of NS binaries (see Eq.~\ref{eqn:binevap}) could become comparable to the segregation time, and therefore the rates of NS-NS mergers could be even lower, if supplied from outside the central region of the nuclear cluster. If more massive stellar-BHs exist, the most massive ones will dominate the inner regions due to strong mass-segregation \citep{alexander09,aharon16}, and can be resupplied into the inner regions faster, enhancing the rates by up to a factor of a few. The larger numbers and faster supply will therefore bias the mass-function of merging BHs through this process to higher masses. Also, the relaxation and mass-segregation times in non-cuspy nuclear clusters, or when no nuclear cluster exists (e.g. for SMBHs more massive than $\sim10^8\,$ M$_\odot$), could be so long that the stellar density around the SMBHs is low. As a consequence, the resupply of stars close to the SMBH cannot be efficiently attained through 2-body relaxation processes, but is more likely to depend on the gas-inflow and in-situ star formation close to the SMBH \citep{ant13,ant14}.

The star-formation rate close to non-resolved regions around SMBHs is difficult to estimate theoretically. Here, we try to use an empirical estimate based on our own resolved Galactic Centre \citep[see e.g.][]{bart09}. Approximately $\sim 200$ O-stars (likely to later form stellar black holes) are observed and inferred to have formed over the last 10 Myrs in the young stellar disk close ($\sim 0.05-0.5$ pc) to the SMBH. The number of lower-mass B-stars in the same environment suggests that similar continuous star-formation has not occurred over the last 100 Myr. Based on these observations we may consider an in-situ formation rate of BHs of $\sim 200/10^8=2\times10^{-6}$ yr$^{-1}$, i.e. comparable to the estimated supply rate of $2.5\times 10^{-6}$ from mass-segregation of BHs from outside the central regions. The comparable formation rate of NSs, however, would increase their resupply rates to the same level as BHs, i.e. much higher than the resupply from NSs migrating in from the outside ($\sim 2\times 10^{-6}$ yr$^{-1}$) and thereby
\begin{equation}
\Gamma_{\rm NS}^{\rm SF}(\msmbh)= 2.8 f_{\rm merge}\ \mathrm{Gpc}^{-3}\ \mathrm{yr}^{-1} \times\left(\frac{4\times 10^6\msun}{\msmbh} \right)^{1/4}\ .
\end{equation}

Table~\ref{tab:rates} reports the resulting rates as a function of the SMBH mass. Upon using the semi-major axis and eccentricity distributions following the prescriptions of \citet{antoper12}, we typically get a merger fraction $\sim 2-5$ times larger than in the case of adopting the initial conditions from \citet{hoan18}. This is probably related to the fact that the semi-major axes of CO binaries are typically smaller in the former case. BH-NS binaries should have mass segregation times similar to BH-BH binaries, hence we use Eq.~\ref{eqn:ratebh}. For all SMBH masses considered in this study, the rates are in the range $\sim 0.17$-$0.52 \ \mathrm{Gpc}^{-3}\ \mathrm{yr}^{-1}$, $\sim 0.06$-$0.10 \ \mathrm{Gpc}^{-3}\ \mathrm{yr}^{-1}$ and $\sim 0.41$-$1.71\times 10^{-3} \ \mathrm{Gpc}^{-3}\ \mathrm{yr}^{-1}$ for BH-BH, BH-NS and NS-NS binaries, respectively. In the star-formation channel, the NS-NS rate may be as high as $\sim 0.04$-$0.16 \ \mathrm{Gpc}^{-3}\ \mathrm{yr}^{-1}$. We note that the merger rate is a decreasing function of the SMBH mass, even though the relative fraction of merger events is typically larger for more massive SMBHs (see Tab.~\ref{tab:models}). On the other hand, more massive SMBHs imply longer relaxation times, that contribute to a reduction in the merger fraction and make the relative rates smaller. Finally, we also run one model where we take $a_{\rm out}^{M}=0.5$ pc, to check how the results depend on this parameter; wide binaries can be affected by LK cycles at distances larger than $\sim 0.1$ pc. In this case, we find that typically our merger fraction $f_{\rm merge}$ is reduced by a factor of $\sim 2$--$3$\footnote{Note that these binaries could however be perturbed by dynamical interactions \citep[see e.g.][]{leigh16}}.

\subsection{Comparison of merger rates to previous studies} 
We find rates comparable with though somewhat different from the ones of  \citet{antoper12} and those of \citet{petr17} for spherical clusters (the latter finds ten times higher rates for the case of a non-spherical nuclear cluster - not modelled, or compared with in our work), but lower with respect to other works that explored the role of the SMBH in reducing the merger timescale of binaries due to Lidov-Kozai oscillations \citep{hoan18}. Other merger channels typically predict larger rates. Typical values for globular clusters are $\sim 2-10$ Gpc$^{-3}$ yr$^{-1}$ \citep{askar17,frak18,rod18} and for nuclear star clusters are $\sim 1-15$ Gpc$^{-3}$ yr$^{-1}$ \citep{ant16}. For reference, the BH-BH merger rate inferred by LIGO is $\sim 12-213$ Gpc$^{-3}$ \citep{abbott17}.

Before directly comparing different estimates, we emphasize that any rate estimate is highly uncertain and should be considered as an order-of-magnitude estimate, since it depends on the specific assumptions regarding the star-formation history in galactic nuclei and the supply rate of compact objects at various distances from the SMBH, which remains poorly constrained in the literature.

We note that previous studies that used a similar approach made use of much larger supply rates, $\sim 10$-$40$ times higher than considered here. The differences arise for several reasons, in particular from the different assumptions on the star-formation rate in the Galactic Centre and nuclear star cluster. \citet{petr17}, \citet{hamer18} and \citet{hoan18} considered a resupply rate of BHs of $10^{-5}-10^{-4}$ yr$^{-1}$ from star-formation, where the latter rate is derived assuming a top-heavy initial mass function from \citep{maness07}. However, this gives rise to several difficulties: (1) This rate is based on the formation rates derived by \citet{lockmann09a} and \citet{lockmann09b} for both NSs and BHs lumped together, while the formation rate for NSs is actually $\sim 8$ times higher than that of BHs for regular (e.g. Kroupa or Miller-Scalo) IMFs, and becomes comparable only for top-heavy IMFs; one can therefore not discuss the number of BHs taking these numbers at face value, but consider the division between BHs and NSs (this issue was accounted for by \citet{petr17}, \citet{hamer18}, but not in \citet{hoan18}), and its dependence on the assumed IMF. (2) The higher rate estimates are based on a top-heavy IMF, which in-turn is based on results from \citet{maness07}. However, this top-heavy IMF was only derived from observations of old low-mass stars and the results are therefore highly problematic for the use in this context, as they are based on a large extrapolation from the low-mass regime up to that of NS and BH progenitors not probed at all by \citet{maness07}. Also note that low-mass stars could already dynamically evolve through mass-segregation and their observed distribution in the GC does not necessarily reflect the actual IMF \citep[e.g. see][]{aharon15}.  Moreover, direct observations of {\emph{massive stars}} in the GC today, though suggestive of a somewhat top-heavy IMF \citep{lu13}, find a much steeper power-law of $-1.7\pm0.2$ compared with $\sim-0.8$ in the Maness et al. study of low-mass stars (which also supersedes the shallow power-law results from from Bartko et al. 2010 regarding massive stars), compared with $-2.3$ for Salpeter or Kroupa IMFs in the relevant mass-range. (3) Even more important, all of these estimates considered star formation throughout the nuclear cluster, rather than the innermost regions, where the induced mergers actually take place (especially in the cases considered by \cite{hoan18}), i.e. the COs would have to migrate inwards over long timescales, and one should then refer to the ex-situ resupply rate discussed above. Indeed, a recent paper by Zhang et al. (2019) exploring the long-term evolution of binaries in the cluster shows that the long-term softening and disruption of binaries due to stars and the SMBH effectively quench the contribution of secular evolution induced mergers, consistent with the points raised above in the case of ex-situ supply.  The overall numbers of COs derived when assuming in-situ formation in these other papers are therefore at least 10 times higher than actually expected in the central parts, and, in fact, are at least 10 times higher than what one can infer from the observed young massive stars in the GC \citep{bart09,bart10}, or those inferred from X-ray sources \citep{muno05}.

We also briefly note that the binary fraction of BHs and NSs in nuclear clusters is not well known, and the different studies make somewhat different assumptions and definitions in their calculations; for example, we calculate the fraction of BHs/NSs given the fraction of massive progenitors in the population and then multiply by the binary fraction, making use of past theoretical and observational studies of BH/NSs, while \cite{petr17} and \cite{hamer18} calculated the total binary fraction, including the dependence on the mass-function derived from stellar evolution. The differences on this point, however, are of order a factor of 1-2, and do not amount to a significant difference. Finally, our models originally considered binaries up to 0.1 pc where most of the current star-formation is observed (the young stellar disk). As mentioned above the merger fractions we find at larger separations up to 0.5 pc are smaller, and therefore, in our models these can contribute at most a comparable number of sources. This potential factor of two in the rate could then also help reconcile some of the differences in comparisons with the previous models which effectively considered star-formation throughout the central pc.

In summary, we believe the overall higher rate estimates near SMBHs by a factor of a few up to $\sim 10s$ in some of the previous studies mostly arise from the different assumptions on the star-formation rate in the Galactic Centre and nuclear star cluster. If one assumes that it formed only through in-situ star formation and that star formation occurred as observed today (at $\lesssim 0.1$ pc) throughout the last $\sim 10$ Gyr, then the resupply rate considered in previous studies should be valid. For the Milky Way, however, these assumptions seem not to be consistent both with the star-formation history over the last $100$ Myrs given the observations of OB stars and X-ray sources (as discussed above), and with the inferred long-term star-formation history over the lifetime of the Galaxy (Nogueras-Lara et al., in prep; R. Sch{\"o}del, private communication), suggesting most of the stars in the Galactic Centre formed very early ($\gtrsim 8$ Gyrs), and at most a few percents formed in the last Gyr. Nevertheless, the overall star-formation history in our Galaxy remains poorly constrained, and one can not exclude that a more vigorous and continuous star-formation could occur in the nuclei of other galaxies, in which case a higher rate estimates may apply, thus a higher merger rate of COs.

\section{Electromagnetic counterparts and observational signatures of SMBH-induced CO mergers}
\label{sect:implications}

As we noted previously (cfr Sec.~\ref{sect:bhnsmergers}) the very high (close to unity) eccentricity, with which the GW signal enters the LIGO band in the scenario explored potentially proviude an important
observational diagnostic of CO mergers induced by LK oscillations.  In the following, we discuss further
observational diagnostics of this merger channel in relation to
possible electromagnetic (EM) counterparts to the mergers.
In particular, mergers of compact object binaries are expected to be associated with a strong release of electromagnetic radiation, if the right conditions arise to power an energetic outflow.

In the case of a NS-NS merger, tidal disruption during the inspiral phase leaves behind an accretion torus surrounding the merged object (either a NS or a BH), unless the two NSs have identical masses \citep{Shibata2006,Rezzolla2010,Giacomazzo2013,Hotokezaka2013,Kiuchi2014,Ruiz2016,Radice2016}. An energetic engine can be driven by rapid accretion onto to the remnant object and/or by dipole radiation losses if the remnant is an hypermassive or stable NS \citep{Giacomazzo2013b,Ciolfi2017}.  Growth and collimation of magnetic fields during the merger, as well as neutrino losses, are then believed to
power a relativistic outflow. Dissipation within the expanding flow, and later interaction of the flow with the interstellar medium, gives rise to radiation that spans a wide window in the electromagnetic spectrum, from high-energy $\gamma$-rays down to the radio.  This basic scenario has been observationally confirmed with the recent event GW170817/GRB170817A
\citep{Abbott2017a}.

Mergers of BH-NS binaries (always resulting in a BH as the resulting compact remnant) are expected to be accompanied by the formation of an hyperaccreting disk only if the mass ratio between the BH and NS does not exceed the value $\sim3-5$, with the precise value depending on the equation of state of the NS and the BH spin \citep{Pannarale2011,Foucart2012,Foucart2018}. For larger mass ratios, the tidal disruption radius of the NS is smaller than the radius of the innermost stable circular orbit, and no disk
will form, resulting in a direct plunge into the BH (see e.g. \citealt{Bartos2013} for a review). If a rapidly accreting disk forms, then the resulting EM phenomenology is expected to be similar to that of the NS-NS case, at least in so far as the bulk properties are concerned.  For the initial conditions explored in this work, $\sim 10$-$20$\% of mergers is expected to have mass ratios $\lesssim 5$, and hence possibly giving rise to an accretion-powered EM counterpart.

In the case of a BH-BH binary merger, there is no tidally disrupted material which can readily supply the accretion power for a relativistic outflow\footnote{Note however that alternative scenarios, involving pure electromagnetic energy, have also been invoked as alternatives to accretion to produce energy 
\citep{Zhang2016,Liebling2016}.}. However, following the tentative detection of a $\gamma$-ray counterpart by the \textit{Fermi} satellite to the event GW150914 \citep{Connaughton2016}, several	ideas were proposed for	providing the merged BH with a baryonic remannt to accrete from \citep{Perna2016,Loeb2016, Woosley2016,Murase2016, Stone2017, Bartos2017,Kimura2017, Janiuk2017,DeMink2017}. Within the context of this study, the scenario proposed by \citet{Bartos2017} is of particular relevance; they note that BH-BH binaries merging within an AGN disk can accrete a significant amount of gas from the disk, well above	the Eddington rate, and	possibly give rise to high-energy EM emission.

Electromagnetic counterparts to binary mergers provide crucial information on the production mechanism of the binaries, since they can potentially allow a much better localization compared to GWs
alone.  A distinctive signature of binary mergers enhanced by LK oscillations in the vicinity of SMBHs is their relatively short merger timescale compared to that of other formation channels. For example, the classical channel of isolated binary evolution predicts merger times $\sim $~100~Myr-15~Gyr \citep{Belczynski2006}. The short lifetimes of the binaries, coupled with their production in the galactic centers, lead to correspondingly short distances traveled prior to mergers.  We find that these distances are typically $\lesssim 0.1$~pc, which makes these merger events practically occurring within the close nuclear region. This constitutes a major	difference with respect to the	standard isolated binary	evolution scenario \citep{Perna2002,Belczynski2006,Oshau2017,Perna2018}: whether it is a small or a large galaxy, the bulk of the merger events occurs at projected distances (from the galaxy center) $\ga 100$~pc\footnote{Note that, even if the isolated binary evolution scenario does predict a fraction of tight binaries with sub-Myr lifetimes \citep{Belczynski2006}, and even ultra-short merger times \citep{Michaely2018}, but the merger sites are still dominated by large scales since isolated binaries are born throughout the galactic disk.}. Localization via EM counterparts hence becomes an especially useful discriminant.

Short GRBs associated with NS-NS mergers, and BH-NS mergers with a small enough mass ratio to allow tidal disruption, are expected to be followed by broadband radiation called afterglow, resulting from the
dissipation of a relativistic shock propagating in the interstellar medium \citep*{Sari1998}. The maximum flux intensity (at any wavelength) is given by
\begin{equation}
F_{\nu,{\rm max}}=
110\;{n_1}^{1/2}{\xi_B}^{1/2}\;E_{52}D_{28}^{-2}\;(1+z)\;{\rm mJy}\;,
\label{eq:Fnu2} \end{equation}
where $E_{52}$ is the explosion energy in units of $10^{52}$~erg, $D_{28}$ the luminosity distance in units of $10^{28}$~cm, $z$ is the redshift, $n_1$ the number density of the interstellar medium in cm$^{-3}$, and it is assumed that the magnetic field energy density in the shock rest frame is a fraction $\xi_B$ of the equipartition value.

The broadband spectrum evolves with time, and we compute it numerically using the formalism of \citet{Sari1998}. For an energy $E=10^{50}$~erg as typical of short GRBs, standard assumptions for the shock parameters and medium ambient density $\sim $ a few cm$^{-3}$ (more typical of the inner regions of a galaxy), the afterglow luminosity in some representative bands (X-rays and radio) at some typical observation times is found to be $L_{[2-10]{\rm kev}}\sim 5\times 10^{45}$~erg~s$^{-1}$ at $t_{\rm obs}=1$~hr and $L_{5GHz}\sim 6\times 10^{30}$~erg~s$^{-1}$~Hz$^{-1}$ at $t_{\rm obs}=7$~days.
In the X-rays, a representative flux threshold is the {\em Swift}/XRT flux sensitivity of $F_{\rm lim}=2.5\times 10^{-13}$~erg~s$^{-1}$~cm$^{-2}$, while in the radio, a 1hr integration with the VLA leads to a flux threshold for detection of $F_{\rm lim}\approx 50\mu$Jy. In both these bands, detection would	
be possible up to a redshift of $\sim~2$, considerably larger than the LIGO horizon\footnote{It should however be noted that in the X-rays, when the shock is still moving at relativistic speeds, relativistic beaming of the emission will lead to a reduced luminosity for jets which are not observed on-axis. The fraction of on-axis jets is expected to be on the order of $1/20$ by taking the jet size of $\sim 16$~deg inferred for short GRBs \citep{Fong2015}.}. The detection distances  are larger than in the isolated binary evolution scenario, for which the large traveled distances lead to a sizable fraction of mergers to occur in low-density environments, where the afterglow luminosity is considerably dimmer.

Additionally note that, independently of the post-merger EM signal, a fraction on the order	of a few $\times 10^{-3}$ of the GW sources is expected to be accompanied by an SN-type precursor \citep{Michaely2018}. This is	due to the fact	that the distribution of the delay time between the
last SN	explosion and the binary merger	has a non-negligible tail of ultra-short times, on the order of 1-100~yr (see also \citep{Dominik2012}).

Detection of an EM counterpart  to a GW event generally allows a redshift measurement. The redshift distribution of the channel studied here would be one that follows the star-formation rate, since the merger times are shorter or at most comparable to the lifetimes of the most massive stars (as a reference, the lifetime of a $\sim 100M_\odot$ star is about 1 Myr). This would hence constitute another observational diagnostics.

The relatively easier prospects for detecting EM counterparts from CO mergers in galactic nuclei makes this channel especially useful for extraction of astrophysical and cosmological information from combined
GW/EM detections. This includes, among other, measurements of the Hubble constant, new tests of the Lorentz invariance, constraints on the speed of GWs, probes of the physics of mergers and jet formation, constraints on the	equation of state of neutron stars \citep[see][and references therein]{Abbott2017a}.

\section{Discussion and summary}
\label{sect:conc}

In this paper, we have revisited the SMBH-induced mergers of compact binaries orbiting within its sphere of influence. While previous studies in the literature adopted the secular approximation for the equations of motion \citep{antoper12,hamer18,hoan18}, here we have performed an extensive statistical study of BH-BH, NS-NS and BH-NS binary mergers by means of $\sim 35000$ direct high-precision regularized $N$-body simulations, including Post-Newtonian (PN) terms up to order PN2.5. 

We have shown that the secular approach breaks down for systems with mild and extreme hierarchies. We used the recent corrections to the maximal eccentricity $e_{\rm max}$ and merger times in the quasi-secular regime \citep[][GPF18]{grish18} and tested it against N-body population synthesis integrations. The total number of mergers is under-predicted by a factor of $\sim 6-10$ in the secular approach, and by a factor of $\sim 2-3$ by the corrected GPF18 model. The CDF of merger times fail to fit either of the distributions, although the D-value distance between the simulated CDF and the GPF18 model is closer than for the secular approach. The difference can be attributed to the original underestimate of $e_{\rm max}$ in the latter model, which leads to slower and less frequent mergers. 

In our numerical simulations, we have considered different SMBH masses, different slopes for the BH mass function and the binary spatial distributions, and different CO binary semi-major axis and eccentricity distributions. We find that the majority of binary mergers happen when the mutual inclination of the binary orbit and its center of mass orbit around the SMBH is $i_0 \sim 90^\circ$, as a consequence of the Lidov-Kozai mechanism. We have also shown that the distributions of the inner and outer semi-major axes of the merging binaries depend mainly on the mass of the SMBH and on the slope $\alpha$ of the binary spatial distribution around the SMBH. On the other hand, the shape of the resulting CO mass distributions depend on the slope $\beta$ of the BH mass function. BH mergers observed by LIGO seem to favour $\beta\sim 1$, if those mergers were to happen around SMBHs. We have also discussed that the fraction of binaries that enter the LIGO band with $e\gtrsim 0.1$ is $\sim 10\%$-$20\%$, larger than previous values found in the literature \citep{antoper12,van16,rand18}, due to the different integration schemes adopted.

We have also calculated the resulting rates as a function of the SMBH mass. We find that the merger rates are a decreasing function of the SMBH mass and are in the ranges $\sim 0.17$-$0.52 \ \mathrm{Gpc}^{-3}\ \mathrm{yr}^{-1}$, $\sim 0.06$-$0.10 \ \mathrm{Gpc}^{-3}\ \mathrm{yr}^{-1}$ and $\sim 0.41$-$1.71\times 10^{-3} \ \mathrm{Gpc}^{-3}\ \mathrm{yr}^{-1}$ for BH-BH, BH-NS and NS-NS binaries, respectively. In the star-formation channel, the NS-NS rate may be as high as $\sim 0.04$-$0.16 \ \mathrm{Gpc}^{-3}\ \mathrm{yr}^{-1}$. We find rates consistent with \citet{antoper12}, but lower with respect to previous works \citep{petr17,hamer18,hoan18}, which may have overestimated the amount of CO binaries supplied to galactic nuclei through star-formation.

We have also discussed the possible EM counterparts of these events. Due to their locations, these mergers may have higher probabilities of being detected also via their EM counterparts, hence making these CO mergers especially valuable for cosmological and astrophysical purposes.

Finally, we note that we have adopted $1$ Myr for the maximum integration time in our simulations, since this limit sets a good compromise between the computational effort and the size of the statistical sample we generate. This choice is further justified by noting that the typical timescale for vector resonant relaxation to operate is $\sim 1$-$10$ Myr, over which the mutual orbital inclination is reoriented by interactions with other background objects, and renders the 3-body approximation insufficient \citep{hamer18}. Also, we have neglected the possible precession of the CO binaries' motion induced by continual weak interactions with other stars and COs in the stellar cusp surrounding the SMBH \citep{alex17}. A comprehensive $N$-body study over long integration timescales that includes both the SMBH-induced Lidov-Kozai oscillations and the detailed effects of the background stars surrounding the CO binaries deserves consideration in future work.

\section*{Acknowledgements}

GF is supported by the Foreign Postdoctoral Fellowship Program of the Israel Academy of Sciences and Humanities. GF also acknowledges support from an Arskin postdoctoral fellowship at the Hebrew University of Jerusalem. EG acknowledges support from the Technion Irwin and Joan Jacobs Excellence Fellowship for outstanding graduate students. EG and HBP acknowledge support by Israel Science Foundation I-CORE grant 1829/12. NL and RP  acknowledge support by NSF award AST-1616157. GF thanks Seppo Mikkola for helpful discussions on the use of the code \textsc{archain}. Simulations were run on the \textit{Astric} cluster at the Hebrew University of Jerusalem. The Center for Computational Astrophysics at the Flatiron Institute is supported by the Simons Foundation.

\bibliographystyle{mn2e}
\bibliography{refs}

\end{document}